\documentclass[12pt]{iopart}
\expandafter\let\csname equation*\endcsname\undefined
\expandafter\let\csname endequation*\endcsname\undefined
\usepackage{amsmath}
\usepackage{hyphenat}
\usepackage{float}
\usepackage{graphicx}
\usepackage{hyperref}
\usepackage{xcolor}

\def\equationautorefname~#1\null{Eq.~(#1)\null}
\def\figureautorefname~#1\null{Fig.~#1\null}
\def\sectionautorefname~#1\null{Sec.~#1\null}
\def\subsectionautorefname~#1\null{Section~#1\null}
\def\subsubsectionautorefname~#1\null{Section~#1\null}

\definecolor{crimson}{HTML}{DC143C}
\definecolor{perfect_green}{HTML}{4FBF26}

\newcommand{\xx}{\ensuremath{\boldsymbol{x}}}

\begin{document}
%[short]{long} titles
\title[Asynchronous and clustered dynamics in noisy inhibitory neural networks]{Coexistence of asynchronous and clustered dynamics in noisy inhibitory neural networks}

%\author{Yannick Feld$^{1,2}$\orcidlink{0000-0003-4305-0430}, Alexander K. Hartmann$^1$\orcidlink{0000-0002-7314-8159}, Alessandro Torcini$^{3,4,5}$\orcidlink{0000-0003-4884-3253}}

\author{Yannick Feld$^{1,2}$, Alexander K. Hartmann$^1$,
Alessandro Torcini$^{3,4,5}$}

\address{$^1$ Université Paris-Saclay, CNRS, CEA, Institut de Physique Théorique, 91191, Gif-sur-Yvette, France}
\address{$^2$ Institut f\"ur Physik, Carl von Ossietzky Universit\"at Oldenburg, 26111 Oldenburg, Germany}
\address{$^3$ Laboratoire de Physique Th\'eorique et Mod\'elisation, CY Cergy Paris Universit\'e, CNRS, UMR 8089,
95302 Cergy-Pontoise cedex, France}
\address{$^4$ CNR - Consiglio Nazionale delle Ricerche - Istituto dei Sistemi Complessi, via Madonna del Piano 10, 50019 Sesto Fiorentino, Italy}
\address{$^5$ INFN Sezione di Firenze, Via Sansone 1, 50019 Sesto Fiorentino, Italy}

\ead{\mailto{alessandro.torcini@cyu.fr}}
%\homepage{https://www.yfeld.de}

\begin{abstract}
A regime of coexistence of asynchronous and clustered dynamics is analyzed for globally coupled homogeneous and heterogeneous inhibitory networks of quadratic integrate-and-fire (QIF) neurons subject to Gaussian noise. 
The analysis is based on accurate extensive simulations and complemented by a mean-field description in terms of low-dimensional {\it next generation} neural mass models for heterogeneously distributed synaptic couplings.
The asynchronous regime is observable at low noise and becomes unstable via a sub-critical Hopf bifurcation at sufficiently large noise. This gives rise to a coexistence region between the asynchronous and the clustered regime. The clustered phase is characterized by population bursts 
in the $\gamma$-range (30-120 Hz), where neurons are split in two equally populated clusters firing in alternation. This clustering behaviour is quite peculiar: despite the global activity being essentially periodic, single neurons 
display switching between the two clusters  due to heterogeneity and/or noise.
\end{abstract}


%\TODO{Global perturbation analysis}
\maketitle
%\TODO{Add alessandros email?}
%\TODO{-ize Endings should be used, not -ise (in cases where both are possible, that is the journal guideline)}
\date{\today}

\section{Introduction}

Since the pioneering studies of Winfree synchronization phenomena in biological populations are usually 
addressed in the context of coupled oscillators \cite{winfree1967}: synchronization is associated
to the emergence of an unique group of oscillators displaying a coherent dynamics \cite{kuramoto1975}.
Besides this phenomenology, one can observe also {\it clustering} phenomena,
where the population breaks down in groups of elements displaying some
sort of coherent evolution \cite{manrubia2004}.

A paradigmatic complex system
where these phenomena have been largely investigated are brain circuits. In this
framework synchronization among a group of neurons can induce the emergences of collective
oscillations (COs) \cite{wang1996, brunel2006}. In this context, the existence 
of inhibitory interactions is fundamental in order 
to promote fast collective oscillatory behaviours in several areas of the brain, in particular in 
the hippocampus and the neocortex \cite{mann2007,bartos2007}. 

 In real systems, and in particular in brain circuits,  noise is unavoidable, therefore
 many analyses have been devoted to its influence on coherent dynamics. In particular,
a common noise source can induce synchronization and clustering 
phenomena, as shown for  globally coupled  or even uncoupled 
limit-cycle oscillators \cite{goldobin2005,nakao2007,gil2010,lai2013}.
This peculiar synchronization induced by common noise is referred in neurophysiology as “reliability” 
\cite{mainen1995}.

But also for the case of independent noise, clustering phenonema have been 
reported for 
heterogenous excitable systems with random coupling strenghts for sufficiently broad distributions 
of the couplings  \cite{sosnovtseva2001}. Furthermore, clustering instabilities have been shown
to affect the synchronized regime in homogeneous inhibitory
networks of spiking neurons subject to additive noise \cite{brunel2006}.

 In this paper, we analyze in details the role of noise in promoting a regime of coexistence among 
clustered and asynchronous dynamics in spiking neural networks. This is a particularly relevant regime, since
brain dynamics in the awake state is typically characterized by an asynchronous activity of the neurons. However, oscillations in the $\gamma$-range (30-120 Hz) can occasionally emerge in relation with information processing, behaviour
and learning \cite{buzsaki2004,susin2021,spyropoulos2022,douchamps2022}.
In particular, we consider an inhibitory spiking neural network 
of quadratic integrate-and-fire (QIF) neurons \cite{ermentrout2008} subject to Gaussian noise. 
The QIF model is quite general, since it represents the normal form 
describing the dynamics of all class I neurons in proximity to a saddle-node on 
a limit cycle bifurcation \cite{ermentrout1986}. Furthermore, for heterogenous QIF networks
exact low-dimensional mean-field (neural mass) models can be derived
in terms of experimentally measurable quantities such as the population firing rate
and the average membrane potential \cite{montbrio2015}. 
Recently, this approach has been extended to encompass extrinsic and endogenous
sources of fluctuations (noise) leading to a hierarchy of low-dimensional neural mass models \cite{goldobin2021}. For his innovation with respect to classical neural mass models (e.g. the Wilson-Cowan one) 
this class of mean-field models has been termed {\it next generation} neural mass models
(for the many possible applications in neural dynamics see \cite{coombes2023}).

We will combine this mean-field analysis with accurate numerical simulations 
\cite{practical_guide2015} to characterize
at a macroscopic and microscopic level the coexisting dynamical regimes, as well as the 
stability of the asynchronous regime and the bifurcations associated to
the emergence of the clustered state. 
To be more specific, the paper is organised as follows. Section 2 is devoted to the introduction of the QIF 
network model and the corresponding mean-field reduction methodology. The macroscopic and microscopic indicators employed to characterize the coherence and regularity of the neuronal dynamics are presented in Subsection 2.3 together with a new stability criterion for the asynchronous state in finite networks inspired by the basin stability analysis \cite{menck2013}.
The linear stability of the neural mass models is analytically evaluated in Section 3 for Gaussian and Lorentzian noise. 
The asynchronous and clustered dynamics are examined in details for heterogenous synaptic couplings in Subsection 4.1 by combining accurate network simulations and neural mass results.
The investigation is extended in Subsection 4.2 to
homogenous couplings but relying only on network simulations.
Spectral analysis of the collective oscillations is reported
in Subsection 4.3 and a brief summary of the obtained results
can be found in Section 5. Finally,  Appendix A reports details on the
numerical simulations, while Appendix B is devoted to the introduction of a criterion 
to select the optimal integration time step in noisy systems.

%%%%%%%%%%%%%%%%%%%%% CONTINUE HERE %%%%%%%%%%%%%%%%%%%%%%%%%%%%%55

\section{Models and indicators}
\subsection{Network Model\label{sec_NetworkModel}}

We consider an inhibitory population of $N$ globally coupled QIF neurons \cite{izhikevich2007},
whose membrane potential evolution is described by the following set of equations 
\begin{align}
     \dot V_i = V_i^2 + \eta_i + \frac{J_i}{N} \sum_{j=1}^N \sum_n \delta(t-t^{(n)}_j)  + \sqrt{2} \sigma \xi_i (t)
\qquad i=1,\dots,N      
    \label{network_model}
\end{align}
where $V_i$ is the membrane potential of the $i$-th neuron,
$J_i < 0$ the inhibitory synaptic coupling strenght and $\eta_i$ the neuronal
excitability. Whenever a membrane potential $V_j$ reaches infinity 
a spike is emitted and $V_j$ is reset to $-\infty$. 
The $n$-th spike-time of neuron $j$ is denoted by $t^{(n)}_j$.

Each neuron is subject to a common synaptic current 
$J_i s(t)$, where 
\begin{equation}
s(t) = \frac{1}{N} \sum_{j=1}^N \sum_n \delta(t-t^{(n)}_j)
\label{activity}
\end{equation}
represents the activity of the network,
as well as to an independent noise term of amplitude $\sqrt{2}  \sigma$,
where $\xi_i(t)$ is a random Gaussian variable with 
$\left< \xi_i(t) \xi_j(0) \right> = \delta_{ij} \delta(t)$.

In the absence of synaptic couplings and of external noise, 
a QIF neuron displays excitable dynamics for $\eta_i < 0$, while
it behaves as an oscillator with period
$T_i =   \pi/\sqrt{\eta_i}$ for positive $\eta_i$. 
For sake of simplicity we will assume homogenous excitabilities,
by fixing $\eta_i = \eta_0 =4.2$, thus all the uncoupled neurons 
will be supra-threshold.

In the following we will consider either heterogeneous 
quenched random couplings following a Lorentzian distribution (LD)
\begin{equation}
h(J_i) = \frac{1}{\pi} \frac{\Delta_J}{(J_i-J_0)^2 + \Delta_J^2}
\label{LD}
\end{equation}
or homogeneous couplings $J_i \equiv J_0$. 
 
In order to characterize the macroscopic behaviour of the network
two indicators will be essential to allow for a comparison with 
the mean-field formulation reported in the next Subsection. 
One is the mean network activity  $s(t)$ \eqref{activity}
and the other
the mean membrane potential, 
defined as follows
\begin{equation}
v(t) = \frac{1}{N} \sum_{i=1}^N V_i(t)  \quad .
\label{eq-mean-membrane-potential}
\end{equation}

The identification of the spike-times 
is subject to some finite thresholding and the numerical
integration of the set of stochastic differential 
equations \autoref{network_model} is explained in details in \ref{Appendix_NumericalIntegration}.
The model is dimensionless, however to report the times (frequencies) 
in physical units, we will assume as a timescale for our dynamics 
$\tau_m = 10$ ms, corresponding to the membrane time constant.

\subsection{Next Generation Neural Mass Model \label{neural_mass_model}}

In the recent years, it has been shown that an exact low-dimensional mean-field formulation
can be developed for fully coupled networks of heterogeneous QIF neurons, with Lorentzian distributed
heterogeneities \cite{luke2013,laing2014,montbrio2015}.
This mean-field (neural mass) model describes the macroscopic
dynamics of the network in the limit $N \to \infty$ in terms of the mean 
membrane potential $v$ \eqref{eq-mean-membrane-potential}
and the population firing rate $r$, which corresponds to the 
network activity $s(t)$ \eqref{activity}. The main assumption
of this approach is that the distribution of 
the membrane potentials is also Lorentzian at any time \cite{montbrio2015}.

This {\it Lorentzian Ansatz} (LA) is  violated if the neurons are randomly connected and/or
in presence of noise. These more general cases can be treated
by introducing a hierarchy of neural mass models taking in account the distortions
to the LD of the membrane potentials \cite{goldobin2021}.
Here we will briefly report the main steps to derive such mean-field formulation in the case
of fully coupled inhibitory network of QIF neurons subject to additive noise. 

In full generality, we can assume that both parameters
$\eta_i$ ($J_i$) are distributed according to a LD $g(\eta)$ ($h(J)$) with 
median $\eta_0$ ($J_0$) and half width at half maximum (HWHM) $\Delta_\eta$ ($\Delta_J$).
In the thermodynamic limit, the network dynamics 
\autoref{network_model} can be characterized in terms of the 
probability density function (PDF) $p(V,t|\boldsymbol{x})$ 
with $\boldsymbol{x}=(\eta,J)$,
which obeys the following Fokker--Planck equation (FPE):
\begin{equation}
\partial_t p(V,t|\boldsymbol{x})+\partial_V\Big[(V^2+I_{\boldsymbol{x}})
p(V,t|\boldsymbol{x})\Big]
 =\sigma^2 \partial^2_V p(V,t|\boldsymbol{x}),
\label{eq102}
\end{equation}
where $I_{\boldsymbol{x}} \equiv \eta + J r(t)$.
In Ref.~\cite{montbrio2015} the authors assumed that 
in the absence of noise the solution of \autoref{eq102} 
converges to a LD  for any initial PDF $p(V,0|\xx )$, i.e., it becomes
\begin{equation}
p(V,t|\boldsymbol{x})=\frac{a_{\xx}}{[\pi(a_{\xx}^2+(V-v_{\xx})^2)]}~,
\label{LDV}
\end{equation}
where $v_{\boldsymbol{x}}$ is the mean membrane potential and  
\begin{equation}
r_{\boldsymbol{x}}(t)=\lim_{V\to\infty}V^2 p(V,t|\boldsymbol{x})=\frac{a_{\boldsymbol{x}}}{\pi}
\end{equation}
is the firing rate for the $\xx$-subpopulation. 
The above LA joined with the
assumption that the parameters $\eta$ and $J$ are independent 
and  Lorentzian distributed
lead to the derivation of exact low-dimensional neural mass models
for globally coupled deterministic QIF networks \cite{montbrio2015}.

Following what was done in \cite{montbrio2015} 
and extending it to noisy systems \cite{goldobin2021}, one can
introduce the characteristic function for $V_{\xx}$,
i.e. the Fourier transform of its PDF, namely
\begin{equation}
\mathcal{F}_{\xx}(k)=\langle{e^{ikV_{\xx}}}\rangle =\mathrm{P.V.}\int\nolimits_{-\infty}^{+\infty}e^{ikV_{\xx}}p(V_{\xx},t|\xx)\mathrm{d}V_{\xx}~,
\end{equation}
where $\mathrm{P.V.}$ indicates the Cauchy principal value.
In this framework the FPE \autoref{eq102} can be rewritten as 
\begin{equation}
\partial_t\mathcal{F}_{\xx}=ik[I_{\xx}\mathcal{F}_{\xx} -\partial_k^2\mathcal{F}_{\xx}]
-\sigma^2 k^2\mathcal{F}_{\xx}\, .
\label{eq103}
\end{equation}
Under the assumption that $\mathcal{F}_{\xx}(k,t)$ is an analytic function of the parameters $\xx$
one can estimate the characteristic function averaged over  the heterogenous population 
$$F(k,t)= \int\mathrm{d}\eta \int\mathrm{d}J  \mathcal{F}_{\xx}(k,t)g(\eta) h(J) $$ 
and via the residue theorem the corresponding FPE, namely
\begin{equation}
\partial_t F=ik\left[H_0 F-\partial_k^2F\right]-|k| D_0 F -\sigma^2 k^2 F\,;
\label{eq104}
\end{equation}
where $H_0 = \eta_0+J_0r$ and $D_0 = \Delta_\eta + \Delta_J r$.

For the logarithm of the characteristic function, ${\Phi(k,t)}=\ln(F(k,t))$, 
one obtains the evolution equation
\begin{equation}
\partial_t\Phi=ik[H_0-\partial_k^2\Phi-(\partial_k\Phi)^2]-|k| D_0 -\sigma^2k^2 \quad .
\label{eq105}
\end{equation}

In this context the LA amounts to $\Phi_L=ikv-a|k|$.
By substituting $\Phi_L$ in \autoref{eq105} for $\sigma=0$ one gets 
\begin{equation}
\dot{v}=H_0+v^2-a^2,
\quad
\dot{a}=2av+D_0\,,
\label{eq:MPR}
\end{equation}
which coincides with the two dimensional exact MF reported in~\cite{montbrio2015}
with $r=a/\pi$. 

In order to consider deviations from the LD, 
the authors of Ref.~\cite{goldobin2021} analysed the following 
general polynomial form for $\Phi$ 
\begin{equation}
\Phi=-a|k|+ ikv- \sum_{n=2}^\infty 
\frac{q_n |k|^n+i p_n |k|^{n-1}k}{n}~.
\label{phi}
\end{equation}
and introduced the notion of complex {\it pseudo-cumulants}, defined as follows
\begin{equation}
W_1\equiv a-iv\,,\quad W_n\equiv q_n+ip_n\,.
\label{eq107}
\end{equation}
By inserting the expansion \autoref{phi} in the \autoref{eq105} one gets
the evolution equations for the pseudo-cumulants, namely:
\begin{equation}
\dot{W}_m=(D_0-iH_0)\delta_{1m}+2 \sigma^2 \delta_{2m} +im\Big(-mW_{m+1}+\sum\nolimits_{n=1}^{m}W_nW_{m+1-n}\Big)~,
\label{eq110}
\end{equation}
where $\delta_{ij}$ is the Kronecker delta function and 
for simplicity we assumed $k>0$. 
It is important to notice that the time-evolution of 
$W_m$ depends only on
the pseudo-cumulants up to the order ${m+1}$, therefore the 
hierarchy of equations can be easily truncated at the $m$-th order 
by setting $W_{m+1} =0$.
As shown in Ref.~\cite{goldobin2021} the modulus of the 
pseudo-cumulants scales as $|W_m| \propto \sigma^{2(m-1)}$ 
with the noise amplitude, therefore it is justified to consider 
an expansion limited to the first few pseudo-cumulants.
 
In this paper, we will consider \eqref{eq110} up to the third order
to obtain the corresponding neural mass model, i.e.
\begin{subequations}
\label{neuralmass}
\begin{align}
%\begin{eqnarray}
    \dot r &=  2 r v + (\Delta_\eta + \Delta_J r + p_2) \pi^{-1} 
%%\nonumber 
\label{neuralmass.1}    
    \\
    \dot v &= \eta_0 + J_0 r - \pi^2 r^2 + v^2 + q_2 
    %%\nonumber 
    \label{neuralmass.2}    
    \\
    \dot q_2 &= 2 \sigma^2 + 4 (p_3 + q_2 v - \pi p_2 r)  
%%    \nonumber 
    \label{neuralmass.3}    
    \\
    \dot p_2 &=  4 ( -q_3 + \pi q_2 r + p_2 v)  
    \label{neuralmass.4}  
%%    \nonumber 
    \\
    \dot q_3 &=   6 (q_3 v - \pi r p_3 - q_2 p_2)
%%    \nonumber  
     \label{neuralmass.5}  
    \\
    \dot p_3 &= 6 (\pi r q_3 + p_3 v) + 3 (q_2^2 -p_2^2)
    \label{neuralmass.6}~,
\end{align}
\end{subequations}    
%%\end{eqnarray}
with the closure $p_4 = q_4 = 0$. 
The macroscopic variables $r$ and $v$ represent the population 
firing rate
and the mean membrane potential, while the terms 
$q_2,p_2,q_3,p_3$ take in account the dynamical
modification of the PDF of the membrane potentials with respect 
to a Lorentzian profile.
Besides the third-order neural mass models, 
we will also consider the second-order one, which can simply be 
obtained by considering Eqs. 
(\ref{neuralmass.1},\ref{neuralmass.2},\ref{neuralmass.3},\ref{neuralmass.4}) by setting $q_3 = p_3 \equiv 0$
 
Since \autoref{neuralmass} is a set of deterministic 
ordinary differential equations,
one can use  standard numerical methods to integrate them. 
In particular, we employed a 4th order Runge-Kutta method \cite{Butcher2008}.
The neural mass results will be compared with network simulations in the following and 
employed to initialize the network in an asynchronous state (see, e.g., \autoref{sec_rho} and \ref{Appendix_ChoosingStepSize}).

\subsection{Indicators}

\subsubsection{Macroscopic Indicators}

The evolution of the membrane potential of a neuron, in particular in the supra-threshold regime, 
can be interpreted as the rotation of the phase of an oscillator and many models have been derived
by employing such an analogy. In terms of these phases the level of synchronization of the 
oscillators (neurons) can be measured in terms of macroscopic order parameters that
we will introduce in the following.
 
For the QIF model, the membrane potential $V_i$ of the $i$-th neuron is usually mapped in the phase 
$\theta_i$ of an oscillator via the following transformation   
\begin{equation}
    \label{phase_trafo}
    \theta_i = 2 \arctan \left( V_i \right)\quad \theta_i \in [-\pi:\pi)~,
\end{equation}
that leads from the QIF network \autoref{network_model} 
to the equivalent $\theta$-neuron network \cite{ermentrout2008}:
\begin{equation}
    \label{theta_model}
  \dot  \theta_i = (1-\cos(\theta_i)) + (1+\cos(\theta_i) \left(\eta_i +  \frac{J_i}{N} \sum_{i=1}^N \sum_n \delta(t-t^{(n)}_j) + \sqrt{2} \sigma \xi_i(t)  \right) \quad .
\end{equation}
Unfortunately, the distribution of the phases $P(\theta_i)$ 
is not uniform even for uncoupled neurons, since
$P(\theta_i) \propto 1/\dot \theta_i$. 
This can therefore lead to apparent synchronization phenomena in 
the $\theta$-space in noisy enviroments \cite{kralemann2007,dolmatova2017}. 

In order to avoid such a problem, the phase $\theta_i$ of the $i$-th neuron 
at a certain time $t$ can be obtained simply by interpolating linearly
between the previous and the next spike time of the considered neuron, as follows  
\begin{align}
    \label{phase_fire}
    \theta_i (t) := 2 \pi \frac{t-t_i^{(n)}}{t_i^{(n+1)}-t_i^{(n)}}-\pi \quad \text{with } t_i^{(n)} \leq t < t_i^{(n+1)}~.
\end{align}
where $t_i^{(n)}$ is the time at which the $n$-th spike is emitted by $i$-th neuron. 

Now that the phases have been defined,  we can introduce suitable order parameters to
measure the level of  phase synchronization in the network
\cite{kuramoto1984, daido1992, strogatz2000, acebron2005, Olmi2016},
in particular we consider the so-called \emph{Kuramoto-Daido order parameters}  
\begin{equation}
    Z_k = z_k e^{i \Psi_k} =  \frac{\sum_{n=1}^N e^{i k \theta_n}}{N}~,
\end{equation}
where $Z_k$ is a complex number 
and $z_k$ and $\Psi_k$ are the corresponding modulus and phase. 
For $k=1$  the usual Kuramoto order parameter \cite{kuramoto1984} is recovered.
For a network of $N$ oscillators one expects 
$z_1 \simeq {\cal  O}(1/\sqrt{N})$ in the asynchronous regime
 and $z_1$ will be finite (one) for a partially (fully) synchronized 
 state.
Unfortunately, $z_1$ is also exactly zero when the oscillators
are equally divided in 2 perfectly synchronized clusters in anti-phase.
To characterize regimes presenting $k$ clusters, Daido \cite{daido1992}
introduced the parameters $Z_k$. 
Indeed, $z_k$ will be one whenever
the system presents $k$ clusters equally 
spaced in phase and equally populated,
while $z_k$ will approach
zero for a sufficiently large network if 
the phases are evenly distributed over the whole interval.

To denote that the order parameters are estimated by employing the phases
defined in terms of the the spiking times as in \autoref{phase_fire} 
we will use a super-script $(s)$, while the lack of a super-script will denote the use of  
the phases defined as in \autoref{phase_trafo}.

In the mean-field framework previously introduced in \autoref{neural_mass_model}, 
$Z_{k}$ can be obtained as follows \cite{Denis}
\begin{equation}
Z_k = z_k e^{i\Psi_k} = \left(1-\frac{W_2}{2}(\partial_{W_1})^2+\frac{W_3}{3}(\partial_{W_1})^3+\frac{W_2^2}{8}(\partial_{W_1})^4+\dots\right)\left(\frac{1-W_1}{1+W_1}\right)^k \enskip .
\end{equation}
Note that the corrections obtained from the higher order pseudo cumulants $W_j$, i.e. with $j>3$, 
should be negligible. In absence of noise and for the usual Kuramoto order parameter $Z_1$ this reduces to the following conformal transformation 
\begin{equation}
    \label{kuramoto_order_eq}
    Z_1 =  \frac{1-W_{1}}{1+W_{1}}~, 
\end{equation}
as shown in Ref.~\cite{montbrio2015}.

Another macroscopic indicator able to distinguish asynchronous regimes
from oscillatory ones characterized by a partial synchrony of the neurons
is the variance $\Sigma_v$ of the mean membrane potential $v$ estimated
over a certain time window $T_W$.
This quantity is expected to be vanishing small $\Sigma_v \simeq {\cal O}(1/\sqrt{N})$ in the asynchronous
regime and finite whenever COs are present.

In order to identify the asynchronous and partially synchronized regime, 
since these are characterized by definitely different values of $\Sigma_v$,
we can define a threshold value $S_\theta$ and whenever
$\Sigma_v < S_\theta$ ($\Sigma_v \geq S_\theta$) the dynamics will be identified as asynchronous (partially synchronized). The threshold value $S_\theta$ is usually chosen as the mean of the values 
measured in the asynchronous and partially synchronized regimes, however 
the identification of the regimes is quite insensitive to the exact value of $S_\theta$.

Also the Kuramoto-Daido order parameter can be employed for this
discrimination, and as we will see in the following for the examined dynamics
the most suitable indicator will be $z_{2}^s$.

\subsubsection{Stability Criterion \label{sec_rho}}

The considered model exhibits in a certain parameter interval
a region of coexistence of two different dynamical regimes:
an asynchronous and a partially synchronous one.
Our goal is to quantify the stability of the asynchronous regime
for the finite network by varying the noise intensity. Therefore, we have introduced 
the following criterion inspired by
the basin stability  criterion \cite{menck2013}, 
which has been applied in many different contexts 
\cite{kim2016, mitra2017, feld19, nauck2022, witthaut2022}.

The main idea is to consider a solution of a system,
perturb this solution several times  with a magnitude that is given by a parameter
and let the dynamics evolve each time. Then one
measures the fraction of how often the system evolves back to a desired
solution. Here,
we proceed as follows: we initialize the values of the membrane potentials $\{ V_i \}$
according to
\begin{equation}
    \label{v_init_eq}
    V_i = \tan \left(\frac{\pi}{2} \frac{(2 i - N - 1)}{N+1}\right) \gamma \Delta_V  + V_0 \quad  i = \{1,2,\text{…}, N\}~,
\end{equation}
with $V_0 = v^*$ and $\Delta_V = \pi r^*$, where $(v^*,r^*)$ are the fixed point solution of
the third-order neural mass model \autoref{neuralmass}. 
Note that for $\gamma=1$ \autoref{v_init_eq} will  result in the Lorentzian distribution that is expected for
an asynchronous regime at equilibrium, while the extreme case $\gamma=0$ fixes all the membrane potentials $V_i \equiv V_0$, i.e.,
it results in a fully synchronized initial state.

For different values of the parameter $\gamma \le 1 $
we simulate the system for a time $T_t$, after which we verify whether the systems is asynchronous or partially
synchronized.  For each value of $\gamma$ we repeat this procedure $M$ times for different
noise realizations and count how many times $M_c$ the system exhibits its asynchronous state after the integration time $T_t$.
Thus, we can measure the stability of the asynchronous state of the chosen configuration via the following indicator  
\begin{equation}
    \label{stab_eq_rho}
    \rho  =  \frac{M_c}{M}\,.
\end{equation}
A completely stable (unstable) asynchronous regime
will correspond to $\rho=1$ ($\rho=0$) for any value of $\gamma$. However, in general 
$\rho$ will be a  function of $\gamma$. In the bistable regime, by decreasing the $\gamma$ value
one will eventually observe a transition towards the partially synchronized regime. 
Thus, the value of $\gamma$ where this ``transition'' happens
is a measure of the stability of the bistable system.

\subsubsection{Characterization of irregular spiking \label{sec_rate_irregular}}

As we will show in the following, the partially synchronized state is characterized by
two clusters of neurons firing in anti-phase. Furthermore, the neurons in each
cluster do not remain in the same cluster over long time periods.
Instead they tend to switch from one cluster to the other,
despite the collective dynamics being always characterized by two clusters 
of neurons firing in alternation. 
Therefore the usual behavior for a neuron is to fire in correspondence with every second neuronal burst, 
but with irregularities in this repetition. 
We want to introduce a measure of these  irregularities in the sequence of spikes of the neurons.

In order to develop this measure we store the sets 
$\mathcal{S}=[s_1=(t_1, i_1), s_2=(t_2, i_2), s_3= \ldots]$ of
firing times and firing neurons in the network for a certain time interval,
where $t_j$ is the firing time and $i_j \in [1,…,N]$, the index of the 
firing neuron. 
Moreover, we also store the bursting times $b_k$ at which the neuronal bursts
occur. These are identified by the maxima in the population firing rate $r$. As a convention we define the burst $k$ as the collection of all the spiking events occurring within the time interval
\begin{equation}
 B_k = \left[\frac{b_{k-1} + b_{k}}{2}, \frac{b_{k} + b_{k+1}}{2}\right)~.
\end{equation}
thus the spike $s_m$ is emitted within the burst $k$ if $t_m\in B_k$.

The regular behavior for a 2 cluster state would be that a neuron, which fires within the burst $k$, would emit its next spike within the burst $k+2$. To analyze the eventual irregularity of the individual neurons we create a 
ordered list $\mathcal{K}_i = \{k_m^{(i)}\}$ reporting the bursts within which
the considered neuron $i$ has fired   
during the observation time interval $T$. The first two bursts $k=0$ and $k=1$ are employed to identify if the neuron belongs to the first
or second cluster,  i.e. $k_{0}^{(i)}= 1$ ($k_{0}^{(i)}= 0$) 
if the spike of neuron $i$ occurred within $B_1$ ($B_0$). 
Then, for each neuron we introduce a counter $E_i$ of ``early spikes''
in the following way:
We go through the list $\mathcal{K}_i$ of bursts to which neuron $i$ has contributed  and we increment $E_i$ by one 
each time $k_{n+1}^{(i)}-k_n^{(i)} < 2$. 
If we had a total of $\mathcal{B}$ bursts in the considered time interval $T$, 
then the fraction of spikes that have been emitted too early by the neuron $i$ is   
\begin{equation}
\lambda^E_i = \frac{E_i}{\mathcal{B}}\,.
\end{equation}
In particular if a neuron would fire during each burst we will have
$\lambda_i^E = 1$.

Similarly, the fraction of ``late delivered spikes'' can be calculated 
by using a second counter $L_i$ that is incremented by one each 
time $k_{n+1}^{(i)} - k_n^{(i)} > 2$, which leads to define the following fraction of late spiking neurons
\begin{equation}
 \lambda^L_i = \frac{L_i}{\mathcal{B}}\,.
\end{equation}
Note that $\lambda^L_i \leq \frac{1}{3}$ due to its definition,
since the counter is incremented whenever $k_{n+1}^{(i)} - k_n^{(i)}$ 
is at least 3.

In summary, the percentage of times the spikes occur 
outside of the expected time-frames, i.e., the 
percentage of irregular spikes, is 
\begin{equation}
    \label{eq_irregularity}
    \lambda_i = \frac{E_i + L_i}{\mathcal{B}} = \lambda^L_i + \lambda^E_i
\end{equation}

The number of neurons for which $\lambda_i=0$ until the time $t$ 
define \emph{the surviving neurons}, i.e., those which fire regularly every 2 bursts as expected.
The fraction of surviving neurons $S(t)$ until the time $t$ can be defined as
\begin{equation}
    S(t) = \frac{\sum_{i \in \mathcal{N}} \delta_{0,\lambda_i(t)}}{\mathcal{N} }~,
    \label{survivor_eq}
\end{equation}
where $\delta$ is the Kronecker delta, and  $\mathcal{N}$ is the number of non-silent neurons.
The silent neurons should be removed from the count, since one always has 
$\lambda_i=0$ for those neurons: a neuron that never spikes obviously 
does not have any associated spike time in an unexpected time interval.
For our analysis we considered $T=55$ s. 

The survival probability $S(t)$ is usually defined as \cite{cox1984} :
$$S(t) = 1 - F(t)  \quad ; $$
where $F(t)=  \int_0^t f(t^\prime) d t^\prime$ is the cumulative distribution function 
and $f(t)$ the PDF of the neuronal survival times, i.e., the time until which the considered 
neuron fires regularly once every two bursts.

\section{Linear stability analysis 
of the asynchronous state \label{linear_analysis}}

In this Section we analyse the stability of the asynchronous
regimes within the neural mass formulation.
For the neural mass model \autoref{neuralmass}, 
the asynchronous states
correspond to fixed point solutions $(r^*, v^*, q_2^*, p_2^*, q_3^*, p_3^*)$. In particular, we will study the stability of these
solutions by considering the linearization of \autoref{neuralmass} in proximity of the considered fixed points, namely
\begin{subequations}
\label{neuralmass_tangent}
\begin{align}
   \delta  \dot r &= 2( v^* \delta r +   r^* \delta v) + (\Delta_J \delta r + \delta p_2) \pi^{-1} 
    \\
    \delta \dot v &=  (J_0  - 2 \pi^2 r^*) \delta r + 2 v^* \delta v + \delta q_2  \\
    \delta \dot q_2 &=  4 (\delta p_3 + q_2^* \delta v +v^* \delta q_2 - \pi p_2^* \delta r -\pi r^* \delta p_2)   \\
    \delta \dot p_2 &=  4 ( -\delta q_3 + \pi q_2^* \delta r + \pi r^* \delta q_2 +  p_2^* \delta v
    +v^* \delta p_2)   \\
    \delta \dot q_3 &=   6 (q_3^* \delta v +v^* \delta q_3 - \pi r^* \delta p_3 -\pi p_3^* \delta r 
    - q_2^* \delta p_2 - p_2^* \delta q_2)  \\
    \delta \dot p_3 &=  6 (\pi r^* \delta q_3 + \pi q_3^* \delta r + p_3^* \delta v + v^* \delta p_3 + q_2^* \delta q_2 - p_2^* \delta p_2)
      \quad .
\end{align}
\end{subequations}

\subsection{Deterministic Case}

Let us start from the case in absence of noise. In this case the mean-field equations
reduce to the exact formulation reported in \cite{montbrio2015}
\begin{equation}
\dot r =  2 r v + (\Delta_\eta + \Delta_J r ) \pi^{-1} \quad \text{and} \quad
    \dot v = \eta_0 + J_0 r - \pi^2 r^2 + v^2~.
    \label{mpr} 
\end{equation}
The fixed point solutions can be obtained by solving the following equations
 \begin{eqnarray}
    v^* &=&  - \frac{\Delta_\eta}{2 \pi r^*} - \frac{\Delta_J}{2 \pi} \label{fp_v} \\
     \pi^2 (r^*)^4 &-& J_0 (r^*)^3 -\left(\eta_0 + \frac{\Delta_J^2}{4 \pi^2}\right) (r^*)^2 - 2 \frac{\Delta_J \Delta_\eta}{4 \pi^2} r^* -
     \frac{\Delta_\eta^2}{4 \pi^2} =0 
    \label{fp_r}    
\end{eqnarray}
for the parameter values considered in this paper, namely inhibitory coupling $J_0=-20$, $\eta_0=4.2$
and $\Delta_J=0.02$ and $0 \le \Delta_\eta \le 0.30$ the system exhibits 2 complex conjugate and 2 real solutions.
Among the real ones only one corresponds to a positive firing rate $r^*$ and is therefore physically acceptable.

The stability of such a solution can be obtained by analysing the linear evolution  
$(\delta r(t), \delta v(t)) = {\rm e}^{\lambda t} (\delta r(0), \delta v(0))$ in proximity of the physical 
fixed point solutions $(r^*,v^*)$. This amounts to solving second order characteristic equations for
the eigenvalue problem associated to \autoref{neuralmass_tangent} with $p_2=q_2=p_3=q_3 = 0$, which gives the following result
\begin{equation}
\lambda_{\pm} = \left(2 v^* + \frac{\Delta_J}{2\pi} \right) \pm \sqrt{2 J_0 r^* -4 \pi^2 (r^*)^2 + \frac{\Delta_J^2}{4\pi^2}}
\quad .
\label{eig}
\end{equation}
For the chosen values of the parameters the square root in \autoref{eig} is always purely imaginary. 
Therefore the fixed point is a focus and, when inserting Eq.~(\ref{fp_v}), 
 the real part of the eigenvalues is simply 
given by
\begin{equation}
{\it Re} \lambda = - \frac{\Delta_\eta}{\pi r^*} - \frac{\Delta_J}{2 \pi} ~.
\label{eig_nonoise}
\end{equation}
The focus is always stable, apart from the fully homogenous situation $\Delta_\eta = \Delta_J = 0$ in which case it becomes marginally stable. The heterogeneities tend to stabilize the focus solutions.
Therefore, even in the case of homogeneous coupling $\Delta_J =0$, a small heterogeneity in the excitabilities measured by $\Delta_\eta$ is sufficient to render the fixed point stable.
This can later be seen for $\sigma=0$ in Fig. \ref{linear_fig_deltaeta}.

\subsection{Gaussian Noise}
\label{gauss}

In presence of additive Gaussian noise of amplitude $\sigma$, we always observe a stable focus
for sufficiently small $\sigma$. The effect of noise is an increase of the value of the firing rate $r^*(\sigma)$ with respect to
the case in absence of noise $r^*(0)$. In particular, 
the correction to the deterministic solution can be written as 
$r^*(\sigma) \simeq r^*(\sigma) + a  \sigma^\alpha$.

\begin{figure}[htb]
    \centering
    \includegraphics[width=0.49\linewidth]{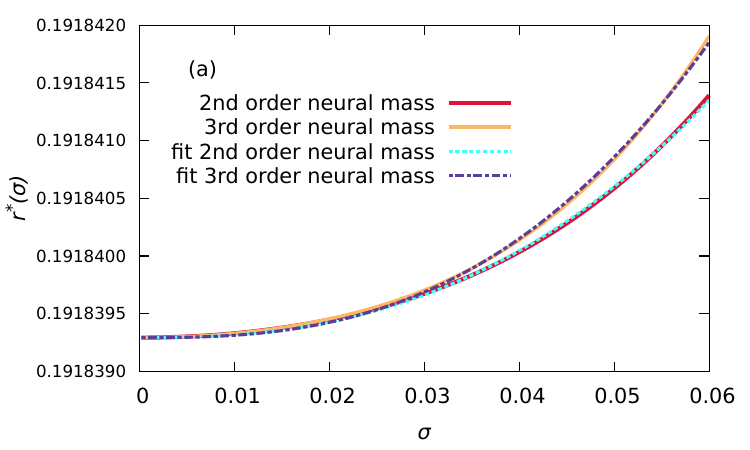}
    \includegraphics[width=0.49\linewidth]{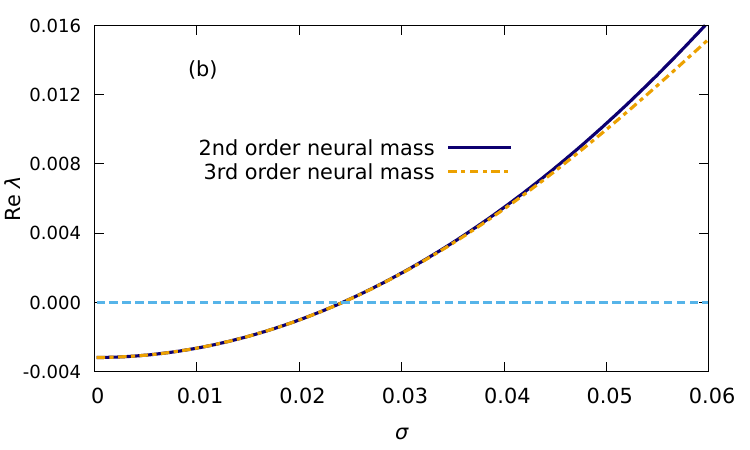}
    \caption{ (a) Fixed point solutions for the firing rate $r^*$ versus the noise
    amplitude $\sigma$ for the 2nd and 3rd-order neural mass models.  Fitting to the data
    with the expression $r(\sigma)=r(0) + a  \sigma^\alpha$ are also reported. (b) 
    Real part of the most unstable eigenvalues for the 2nd and 3rd-order neural mass model     versus $\sigma$. Parameters are set to $\eta_0=4.2$, $J_0=-20.0$, $\Delta_\eta = 0$, and $\Delta_J=0.02$.
      \label{linear_fig}}
\end{figure}

For $\Delta_\eta=0$ and the parameters
usually employed in this analysis 
one obtains $\alpha \simeq 2.5$ for the 2nd-order neural mass model,
 while the growth is even faster for the third-order model
 with $\alpha \simeq 2.7$, as evident from  \autoref{linear_fig} (a).
 
To analyze the stability of the fixed points
we have identified the corresponding eigenvalues 
by solving the associated characteristic polynomial,
that can be of the 4th (6th) order depending if we consider the neural mass model to the 2nd (3rd) order. 
The linear stability analysis reveals that the 4 eigenvalues for the 2nd order neural mass
are two complex conjugate pairs, whose real parts are definitely negative for $\sigma=0$
and $\Delta_\eta$ and/or $\Delta_J$ not zero, as evident from \autoref{eig_nonoise}.

\begin{figure}[htb]
    \centering
    \includegraphics[width=0.6\linewidth]{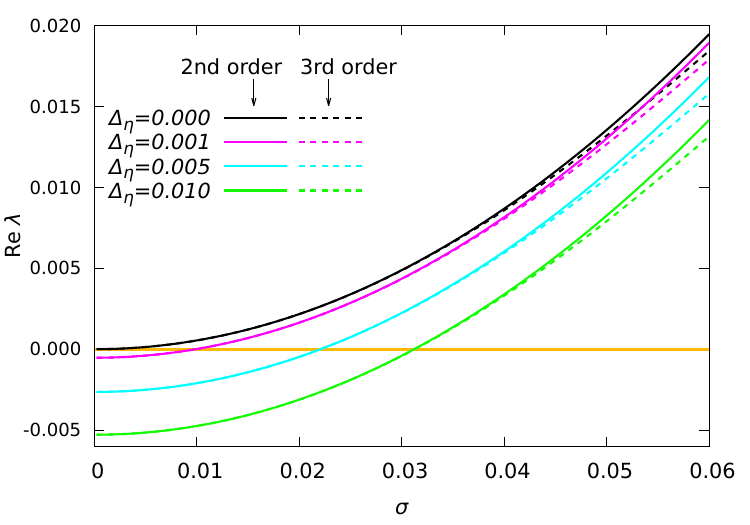}
    \caption{Real part of the most unstable eigenvalues for the 2nd and 3rd order neural mass model versus $\sigma$ for different level of heterogeneity $\Delta_\eta$. Parameters are set to $\eta_0=4.2$, $J_0=-20.0$ and $\Delta_J = 0$.
         \label{linear_fig_deltaeta}}
\end{figure}

As shown in Fig.  \ref{linear_fig} (b),
noise  destabilizes the fixed point, since it leads to
an increase of the real part of the maximal eigenvalues.
In particular, these two eigenvalues can cross the zero axis at a critical noise amplitude $\sigma_{H}$.
Thus indicating that the fixed point solution becomes
unstable via  a Hopf bifurcation giving raise to COs.
For the case shown in \autoref{linear_fig} (b) we have $\sigma_H \simeq 0.0243$. The third order model displays 3 pairs of complex conjugates eigenvalues, however the fixed point looses stability exactly at the same $\sigma_H$ value via a Hopf bifurcation (see \autoref{linear_fig} (b)).
The effect of noise on the stability of the foci is analogous
for a network with homogenous couplings ($\Delta_J=0$) and heterogeneous currents $\Delta_\eta >0$, as shown in \autoref{linear_fig_deltaeta}.
As we will see in the following these Hopf bifurcations are sub-critical, 
thus leading to a coexistence region between asynchronous dynamics and COs.

\subsection{Lorentzian Noise}

It is worth mentioning in this context that, 
by assuming that the white noise terms $\xi_i(t)$ are
Lorentzian distributed, it is still possible to obtain the corresponding low-dimensional neural mass model in an exact manner \cite{cestnik2022,clusella2024}. In particular, by assuming that
the $\xi_i(t)$ random term follow a LD centered in zero and with HWHM $\Gamma$, one can obtain the following two-dimensional neural mass model \cite{clusella2024}
\begin{equation}
\dot r =  2 r v + (\Delta_\eta + \Gamma + \Delta_J r ) \pi^{-1} \quad \text{and} \quad
    \dot v = \eta_0 + J_0 r - \pi^2 r^2 + v^2~,
    \label{mpr_lornoise} 
\end{equation}
which is identical to \autoref{mpr} apart for the $\Gamma$ term that contributes exactly as the HWHM $\Delta_\eta$ of the neural excitabilities $\{ \eta_i \}$ to the mean-field dynamics. Therefore, in the thermodynamic limit the Lorentzian noise
can be assimilated to a quenched disorder in the 
heterogeneities as shown also in Refs.~\cite{cestnik2022,clusella2024}. In the present context, this implies that the neural mass model
\autoref{mpr_lornoise} displays only stable foci 
solutions as \autoref{mpr} and no collective oscillations
are observable, contrary to the case where  the noise is Gaussian distributed.

\section{Numerical results}

In this Section we will analyze and characterize the clustering transition induced by
noise. In particular, we will first investigate heterogenous couplings, 
where we fix $\Delta_J=0.02$, in this case we will compare the results obtained within
the mean-field approach with network simulations. Successively, we will examine the homogenous
situation, where $\Delta_J=0$, by relying only on network simulations. If not specified otherwise we fix the parameters to the
following values $J_0=-20$, $\eta_0=4.2$ and $\Delta_\eta = 0$, and we consider an inhibitory network of size $N=200000$  subject to Gaussian additive noise.

\subsection{Heterogeneous synaptic couplings}

In Subsection \ref{gauss} we have shown that in the mean-field formulation the asynchrous regime remains stable up to a noise of amplitude $\sigma_H \simeq 0.0243$, where it destabilizes via a Hopf bifurcation.
In this Subsection we will characterize such a transition and the observed regimes in full details.

\subsubsection{The clustering transition}

In order to understand if the transition is super- or sub-critical, we perform simulations by varying quasi-adiabatically the noise amplitude $\sigma$ (for details see \ref{Appendix_NumericalIntegration})
and by measuring for each value of $\sigma$ the variance $\Sigma_v$ of the mean membrane potential.

\begin{figure}[htb]
    \centering
    \includegraphics[width=0.6\linewidth]{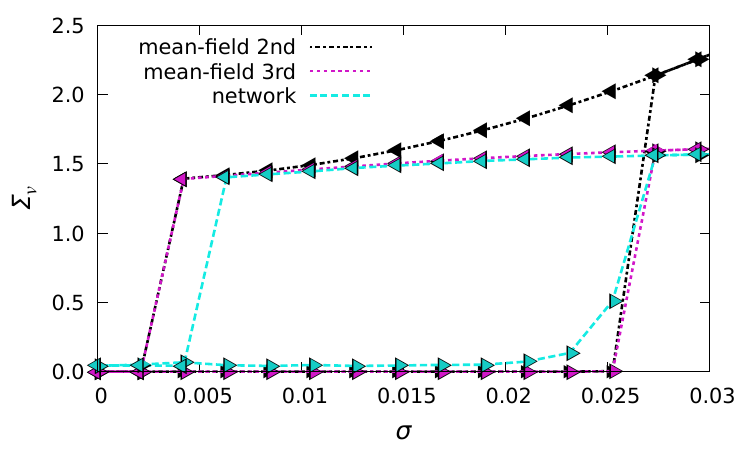}
    \caption{Variance $\Sigma_v$ of the mean membrane potential $v$ versus noise amplitude $\sigma$
    obtained via quasi-adiabatic simulations. The decrease or increase of $\sigma$ performed during the 
    adiabatic simulations is indicated by the direction of the triangles' tip. 
    The dashed lines are meant for visual aid.
    The parameters for the quasi-adiabatic simulations are $\Delta \sigma = \frac{0.04}{19}$, $t_T = 20$ s,
    $t_S = 25$ s, for the neural mass we employed $t_T = 100$ s to allow the system to better relax, and
    the network size was fixed to $N=200000$. Other parameters as in  Fig.  \ref{linear_fig}.
    \label{adiabatic_hetero_fig}}
\end{figure}

As shown in \autoref{adiabatic_hetero_fig}, for the 2nd and 3rd order neural mass models 
we observe that starting from $\sigma=0$ and by increasing the noise amplitude, the variance $\Sigma_v$ remains zero until a value near $\sigma_H$ is reached, 
as expected for a constant 
mean membrane potential ($v=v^*$).  Afterwards it jumps to some finite value
due to the emergence of COs. Once
the noise amplitude reaches $\sigma=0.03$ the quasi-adiabatic 
simulations are then continued by decreasing $\sigma$ 
in steps of $\Delta \sigma$.
In this case $\Sigma_v$ stays finite down to values  
$\sigma_{SN} \simeq 0.004$ and then
at even smaller value of $\sigma$ returns to zero. 
This scenario is typical for a sub-critical Hopf
bifurcation, characterized by the coexistence of oscillatory and stationary behaviours
in the range $\sigma \in [\sigma_{SN},\sigma_H]$. In particular, at $\sigma_{SN}$ one
expects that the oscillatory solution will disappear via a saddle-node bifurcation of limit cycles.

The network simulations agree quite well with those of the 3rd order neural mass model,
apart some finite-size effects that imply finite values ${\cal O}(1/\sqrt{N})$ for $\Sigma_v$
even in the asynchronous regime and a backward transition from the oscillatory to the
asynchronous regime occurring at a larger noise amplitude, namely $\sigma \simeq 0.006$, 
instead that at $\sigma_{SN}$.
In contrast, the 2nd order neural mass model displays  clear differences
with the 3rd order one and 
the network simulations in the oscillatory regime for $\sigma > 0.01$ 
(see black triangles in Fig. \ref{adiabatic_hetero_fig}).
This is probably due to an instability of the 2nd order model at large noise amplitudes.

\begin{figure*}
    \centering
       \includegraphics[width=\linewidth]{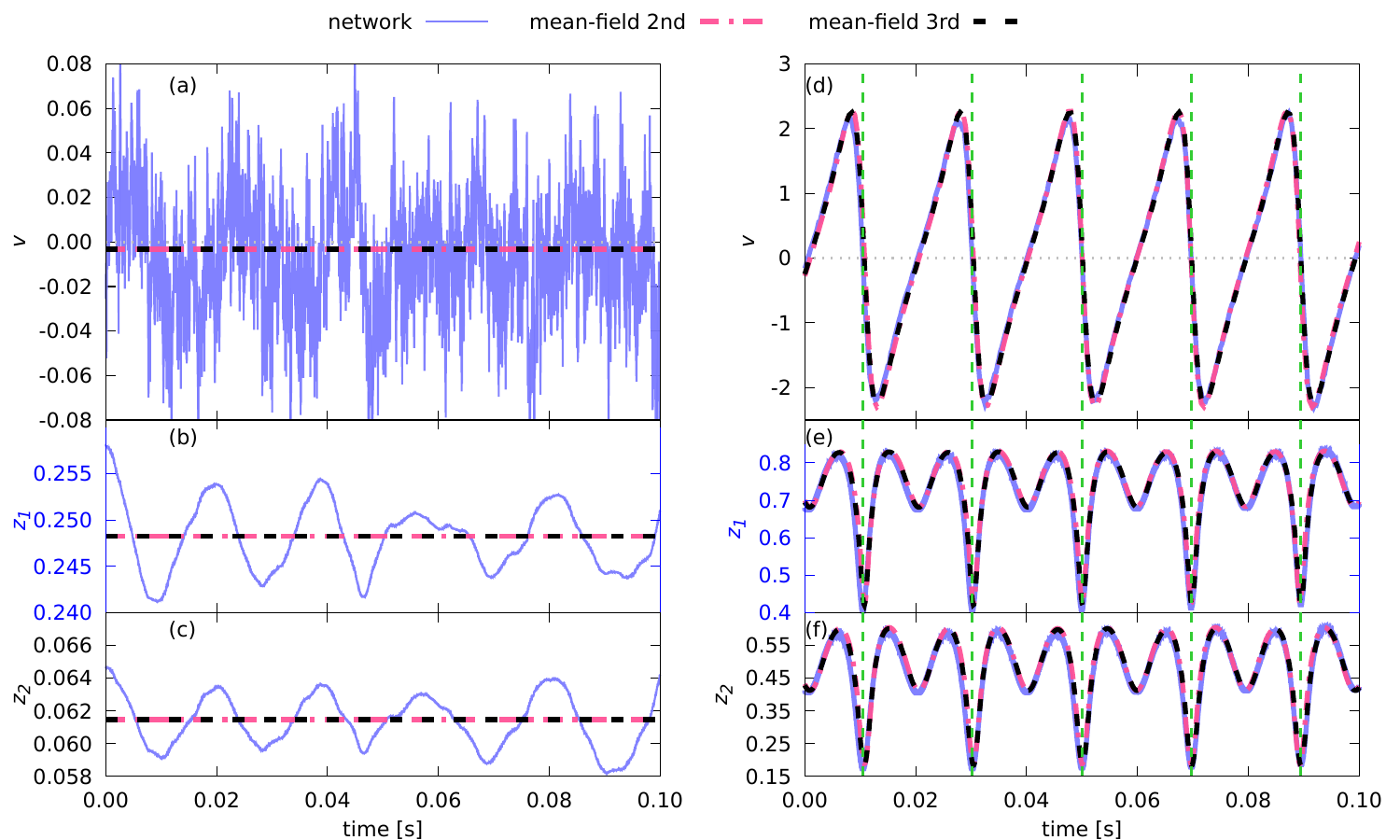}
    \caption{The mean membrane potential $v(t)$ (a and d) and the Kuramoto-Daido order parameters $z_1$ (b and e)
    and $z_2$ (c and f) versus time.
    The data refer to the coexistence regime, namely $\sigma=0.00842$, the asynchronous (clustered) state are reported on the left (right).  Other parameters as in  Fig.  \ref{linear_fig}. 
    \label{fig_vKurmatoCmp}
    }
\end{figure*}

\subsubsection{Coexisting regimes}

Let us now examine the macroscopic properties of the asynchronous and clustered regimes in the
coexistence region in more detail. In order to gain some insight we report 
the mean membrane potential $v$ versus time for the two regimes at $\sigma=0.00842$ in Fig. \ref{fig_vKurmatoCmp} (a and d). 
In the asynchronous state $v$ is exactly constant for the neural mass simulations,
while it displays small erratic fluctuations when obtained from network simulations. 
This is due to the fact that the stable fixed point is a focus, therefore
the presence of finite-size fluctuations excites continuosly relaxation oscillations towards the
focus. As shown in \autoref{fig_vKurmatoCmp} (d)  $v(t)$ is periodically oscillating in the clustered regime
and in this case the network simulations agree quite well with the neural mass results
obtained for both 2nd and 3rd order models.

It is interesting to examine the level of synchronization in the two regimes
as measured by the Kuramoto order paramters $z_1$ and $z_2$, see \autoref{fig_vKurmatoCmp} (b,c and e,f). 
In the asynchronous state shown in panels (b,c) the neural mass results give a 
finite value for $z_1$ and $z_2$,
while for an asynchronous regime one would expect zero values in the mean-field limit.
The values of $z_1$ and $z_2$ obtained by the network simulations oscillate 
in an irregular fashion slightly around the mean-field value. For what concerns the clustered regime, the order 
parameters reveals periodic oscillations  with the same period as $v(t)$ and
significant amplitudes. In this case the 
neural mass and the network results essentially coincide as shown in panels (e and f).

\begin{figure}[htb]
    \centering
    \includegraphics[width=0.8\linewidth]{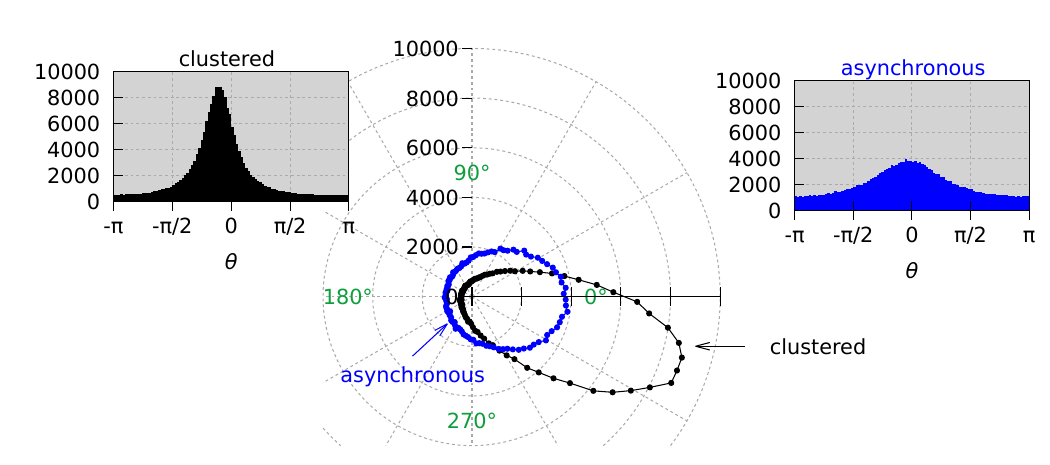}
    \caption{ Snapshots of the phases as calculated via the expression \eqref{phase_trafo} 
    displayed as a polar plot for asynchronous (blue) and clustered (black) regimes. 
    The insets show the histograms of the same data done over 100 bins.      
    The data refer to a system size $N=200000$ and $\sigma=0.00842$.  Other parameters as in  Fig.  \ref{linear_fig}.
    \label{fig_polar_async}}
\end{figure}

In order to understand the reason why $z_1$ and $z_2$ have a finite value in the asynchronous regime
let us investigate the distribution of the phases as defined in \autoref{phase_trafo}.
Histograms of these phases are shown in \autoref{fig_polar_async} for the asynchronous and clustered regime.
In the asynchronous regime the phases are not equally distributed 
in $[-\pi;+\pi]$ as expected they exhibit a peak around zero instead.
This peak is much more pronounced in the clustered regime, but there is no evidence of
the two clusters. This is due to the fact that the phase definition
\autoref{phase_trafo} is related to the membrane potential value, whose values
also displays similar unimodal PDFs 
(see the Supplementary Material for animations of the phase histograms from \autoref{fig_polar_async} 
and the corresponding membrane potential histograms), and not to the
firing time of the corresponding neuron, thus making this phase unsuitable to characterize
the observed neural dynamics.

 Let us now consider the distributions of the phases as 
 obtained by the
 firing times via the definition \autoref{phase_fire} for the asynchronous and
 oscillatory regimes. The results are shown in \autoref{fig_polar_both}.
 Note that animations of these histograms can be found in the Supplementary Material.
 In the asynchronous case, as expected, the phases are uniformly distributed.
 The results for the oscillatory regimes reveal that the neurons are arranged in two clusters
 in phase opposition (at a distance $\pi$) from one another. In this case we expect that 
 the Kuramoto order parameter $z^s_1$ ($z_2^s$) should be zero (order one)
 since the 2 clusters are in phase opposition.

\begin{figure}[htb]
    \centering
    \includegraphics[width=0.8\linewidth]{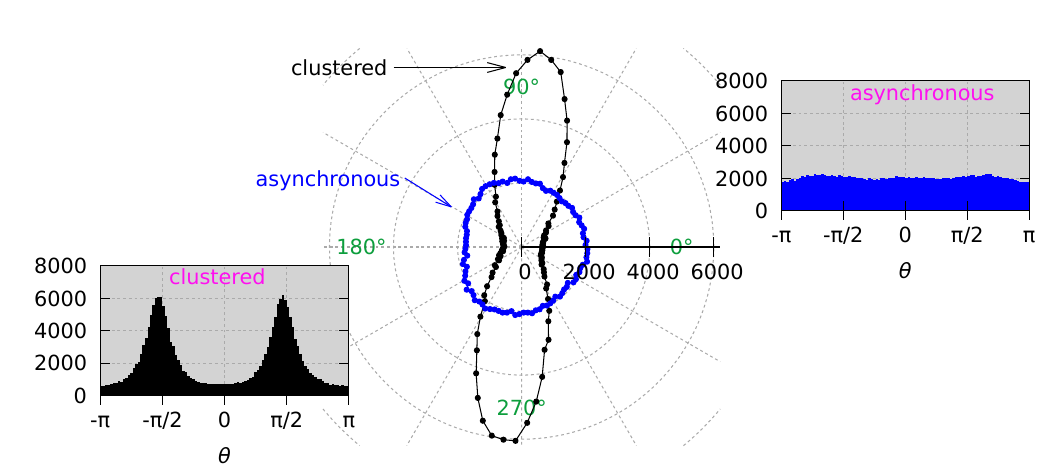}%data/hetero/polar_both/polar_vortrag.pdf
    \caption{
    Snapshots of the phases as calculated via the expression \eqref{phase_fire} 
    displayed as a polar plot for asynchronous (blue) and clustered (black) regimes. 
    The insets show the histograms of the same data done over 100 bins.
    The data refer to a system size $N=200000$ and $\sigma=0.00842$. Other parameters as in  Fig.  \ref{linear_fig}. 
    }
    \label{fig_polar_both}
\end{figure}

\begin{figure}[htb]
    \centering
    \includegraphics[width=0.49\linewidth,scale=0.25]{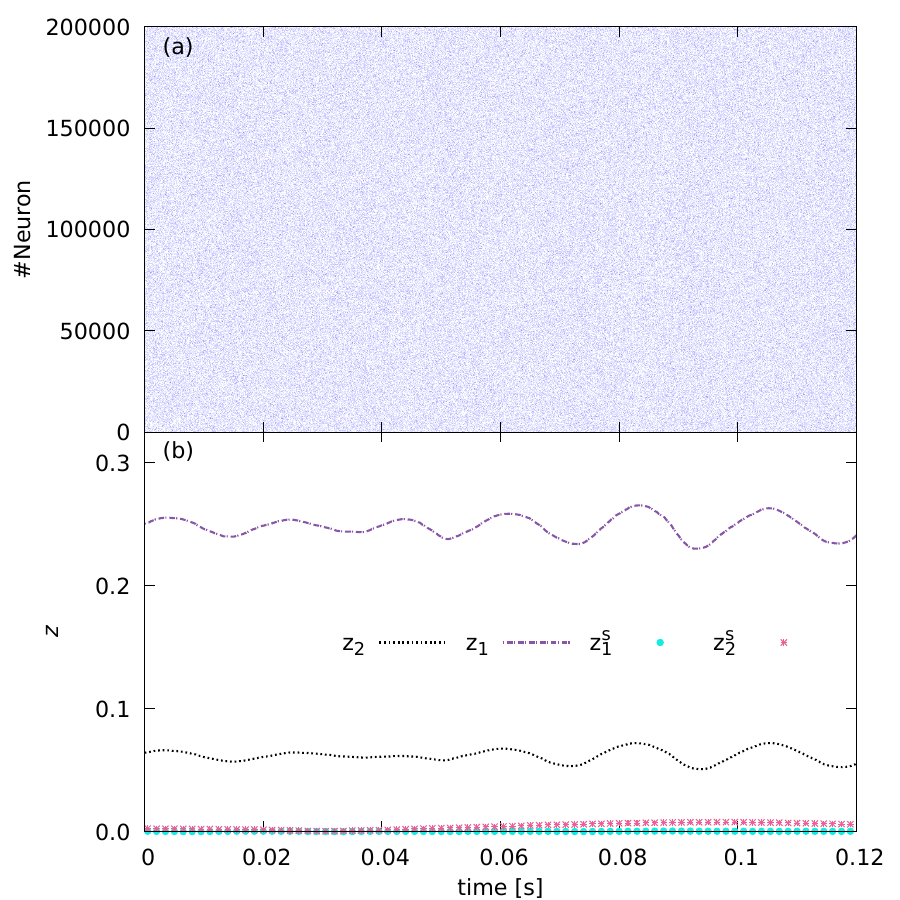}%data/hetero/raster_kuramoto_both/async/raster_and_kur_async_missing_png.pdf
    \includegraphics[width=0.49\linewidth,scale=0.25]{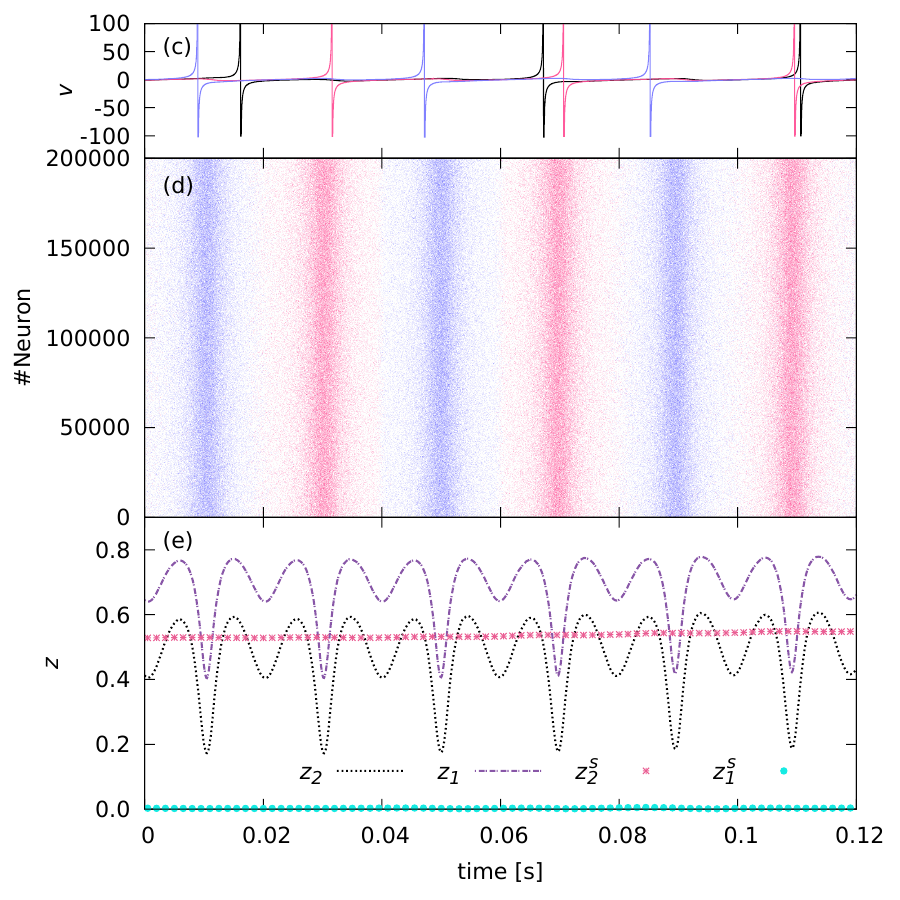}%data/hetero/raster_kuramoto_both/raster_and_kur_sync_missing_png.pdf
    \caption{Raster plot (a and d) and the corresponding order parameters $z_k$ and $z_k^s$ (b and e) versus time. 
    In (c) the membrane potential of three generic neurons are displayed. 
    The results for the asynchronous (clustered) regime  are shown in the left (right) row.  
    In the raster plot (d) the color of the dots indicates in
    which cluster the corresponding neuron is at time $t=0$ based on their next spiking event. 
     The data refer to a system size $N=200000$ and $\sigma=0.00842$. Other parameters as in  Fig.  \ref{linear_fig}.  
      }
    \label{fig_raster_async_hetero}
\end{figure}

To get some more insight on these two dynamical states, we will examine the
raster plots  as a measure of the microscopic network activity joined to
the traces of the Kuramoto-Daido order parameters $z_k$ and $z_k^s$ for the macroscopic
counterpart. These are shown in \autoref{fig_raster_async_hetero}.

In the asynchronous regime, the raster plot in panel (a) does not display any structure 
and the corresponding Kuramoto-Daido order parameters $z_k^s,~k=1,2$ estimated
by the firing-times are of ${\cal O}(1/\sqrt{N})$ as expected (see panel (b)). 
As shown in panel (b) the values of $z_{1}$ and $z_{2}$ are instead definitely finite due to the fact that the phases obtained via the transformation \autoref{phase_trafo} are not uniformly distributed, even in this regime.

 In the clustered regime the raster plot (shown in panel (d)) 
 reveals bursts of activity of the neurons
 interrupted by a low activity phase.
 In each population burst roughly 50\% 
 of the neurons participate.
 In the raster plot, the spiking times are visualized by red and blue colored dots 
 based  on which cluster the corresponding neuron belonged at time $t=0$,
 for which we used the next spiking event of the corresponding neuron.
 In the time window reported in panel (d) the clusters are apparently stable, however on a longer run
 the two ensembles will mix up completely despite the macroscopic dynamics remaining always characterized
 by two equally populated clusters. To exemplify these behaviours in panel (c) the membrane potential traces 
 for 3 characteristic neurons have  been reported: the red (blue) neuron is always firing within the red (blue) burst, while the black one is initially firing within the blue burst but then it skips 2 population bursts and finally joins the red burst. 

 In panel (e) we report the corresponding Kuramoto-Daido order parameters versus time,
the parameters  $z_{1}$ and $z_{2}$ display a periodic behaviour and attain the minimum value
whenever a burst occurs, due to the repulsive nature of the couplings. On the other hand,
$z_1^s$ stays always close to zero as expected for two phase clusters in phase opposition,
while $z_2$ has a constant finite value larger than 0.5 indicating that the composition of the clusters is stable
in time.

\subsubsection{Stability of the asynchronous regime}
\label{stab_asynch}

In this Paragraph we will analyze the stability of the asynchronous
state for increasing noise amplitude. For this we will employ
the stability indicator $\rho = \rho(\gamma)$ 
introduced in \autoref{stab_eq_rho}
as a function of the parameter $\gamma$  controlling the
initial distribution of the membrane potentials according to \autoref{v_init_eq}.
The value $\gamma=1$ ($\gamma=0$) corresponds to an initialization of the 
neurons with membrane potentials distributed according to the LD 
expected for the asynchronous case (with identical values of the membrane potentials $V_0$).
 
\begin{figure}[htb]
    \centering
    \includegraphics[width=0.7\linewidth]{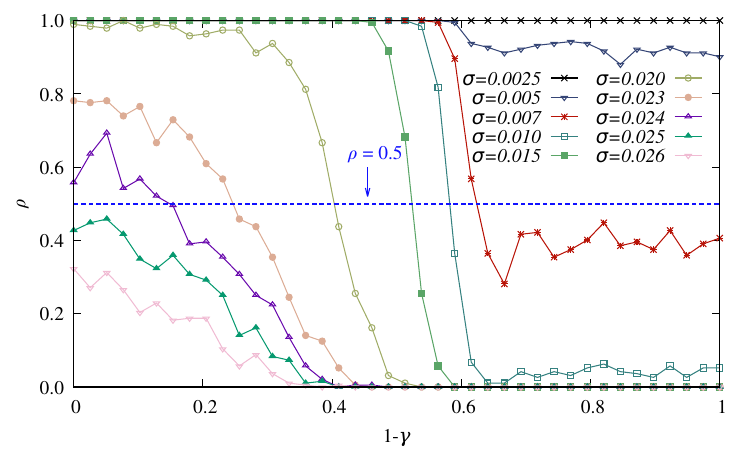}%data/hetero/stability_scan/hetero_stability_scan.pdf
    \caption{Stability indicator $\rho$ versus $1-\gamma$, where $\gamma$ is the factor entering in \eqref{v_init_eq}. 
    For each measurement of $\rho$ we considered $M=192$ different realizations  and
    each time we estimated  $\Sigma_v$ from a time series of 
    $T_W=0.3$ s after discarding a transient $T_t = 16$ s to evaluate if the system remains asynchronous or becomes clustered.   
    The dashed line indicates $\rho=0.5$, i.e., where the system has an equal probability
    to move towards either the asynchronous or the clustered state.
      The data refer to a system size $N=32000$, other parameters as in  Fig.  \ref{linear_fig}.    
    }
    \label{fig_stab_hetero}
\end{figure}

We have estimated the stability indicator $\rho(\lambda)$ for a system size
$N=32000$ (due to cpu limits) and the results are reported in 
\autoref{fig_stab_hetero} for different noise amplitudes.
For the noise amplitude $\sigma=0.0025$, which is smaller 
than $\sigma_{SN} \simeq 0.004$ and therefore 
outside the coexisting region, we observe that  
the asynchronous state is stable for any $\gamma$-values, as expected.
For larger noise amplitude $\sigma > \sigma_{SN}$, large perturbation of  the
asynchronous distribution, as measured by $1-\gamma$, can induce transitions
towards the clustered regime with some finite probability.

At noise amplitudes $\sigma \ge 0.015$, 
for sufficiently synchronized initial conditions, 
namely $\gamma < 0.4$, the system has a probability of almost $100\%$ 
to leave the asynchronous state, thus indicating a clear coexistence of the 2 regimes.
For $\sigma \ge 0.023$ (i.e. in proximity of the Hopf bifurcation identified in the mean-field formulation $\sigma_H = 0.0243$) the probability to stay in the asynchronous case is smaller than 100\% even for the unperturbed initial conditions, corresponding
to $\gamma=1$, in this case we expect that by simulating for a longer time period $T_t$ we would actually measure $\rho=0$. 

In order to identify the critical noise amplitude above which the asynchronous state is 
unstable, we measure  the value ${\bar \gamma}$ for which 
$\rho({\bar \gamma})$ crosses $\frac{1}{2}$ for various noise amplitudes $\sigma$. Thus, this indicates that for $\gamma = {\bar \gamma}$ one has 50\%
of probability to end in the asynchronous or in the clustered state.
For this we estimated the indicator with more precision in proximity of $\rho({\bar \gamma}) \approx \frac{1}{2}$
To this end we considered $M=384$ realizations of the initial perturbed state for each value of $\gamma$,
and we estimated via an interpolation the value ${\bar \gamma}$. In particular, we expect that
the standard deviation of $\rho(\gamma)$ will be maximal at exactly one half, therefore we fitted
such standard deviation to a Gaussian curve for different $\gamma$ value and we extrapolated
the value $\bar \gamma$ where the curve attains its maximum.  

\begin{figure}[htb]
    \centering
    \includegraphics[width=0.6\linewidth]{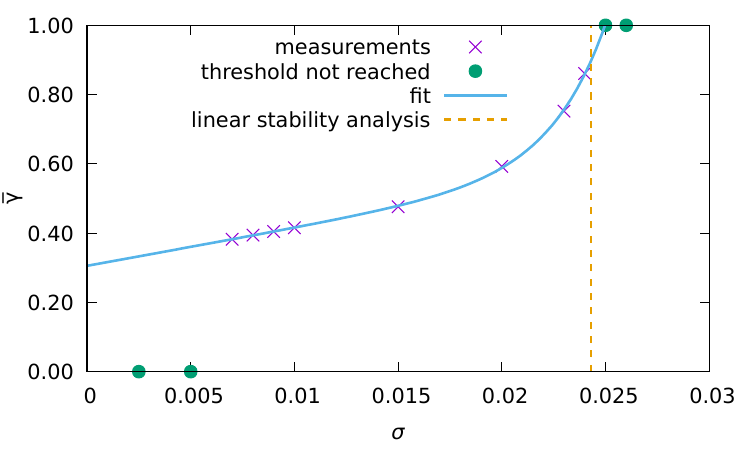}%data/hetero/stability_transition/hetero_transition.pdf
    \caption{ ${\bar \gamma}$ versus the noise amplitude $\sigma$.      
    The green circles represent $\sigma$ values for which $\rho(\sigma)=0.5$ was 
    never reached. The blue solid line is a fit to the data (violet crosses) performed via
    the expression \eqref{fit-stab-pos}, the fit is employed to extrapolate 
    the critical noise value $\sigma_{c}$ for which $\rho(\sigma_{c}, {\bar \gamma}=1)=0.5$.
    The vertical dashed line shows the mean-field prediction $\sigma_H$ for the onset of COs.
    Errorbars are omitted since the errors are smaller than symbol size.  Other parameters as in Fig. \ref{fig_stab_hetero}
    apart the number of different realizations, that now is $M=384$.   
        }
    \label{fig_trans_fit_hetero}
\end{figure}

In \autoref{fig_trans_fit_hetero} we show the obtained values ${\bar \gamma}$ as a function of noise amplitude
$\sigma$ (violet crosses),
we observe that ${\bar \gamma}$ grows with the noise amplitude and approaches the value ${\bar \gamma}=1$.
To identify the critical noise amplitude $\sigma_c$ above which the system always ends up
in the clustered regime for any $\gamma$ value we have fitted the numerical data with this function:
 \begin{equation}
    \label{fit-stab-pos}
    f(\sigma) = 1 + a \left(1- e^{b(\sigma-\sigma_c)} \right)+ c (\sigma-\sigma_c)~.
\end{equation}
By excluding from the fit the data where the threshold value $\rho=0.5$ was not reached 
(green points) we obtain the following parameter values
  $a=-0.42(1)$, $b=376(28)$, $c=10.9(8)$  and $\sigma_c=0.02498(6)$.
  As you can see the fit works pretty well.
  The extrapolated critical value of the noise is in quite good agreement with the mean-field
  result obtained from the linear stability analysis of the asynchronous state that indeed was $\sigma_H = 0.0243$,
  the difference on the third significative digit can be due to finite-size and nonlinear effects.

  In summary the new method here introduced to study the stability of the asynchronous regime works
  reasonably well when compared with the linear stability analysis, that in the present case is feasible due 
  to the existence of low-dimensional mean-field formulations, but usually in a high dimensional network is quite
  difficult to implement. Therefore, this new method can represent an useful alternative to the linear stability analysis
  and it can  find applications in many complex network systems.  Furthermore, it gives
also information concerning the basins of attraction of the two regimes in the coexistence region,
that the linear stability analysis is unable to provide.

\subsubsection{Characterization of the clustered dynamics}

In this Paragraph we would like to examine the clustered
dynamics of the neurons in more details. In particular, we wish to characterize the erratic behaviours
that lead the neurons to deviate from a perfectly locked evolution,
where the neurons fire every second burst.

\begin{figure}[htb]
    \centering
    \includegraphics[width=0.6\linewidth]{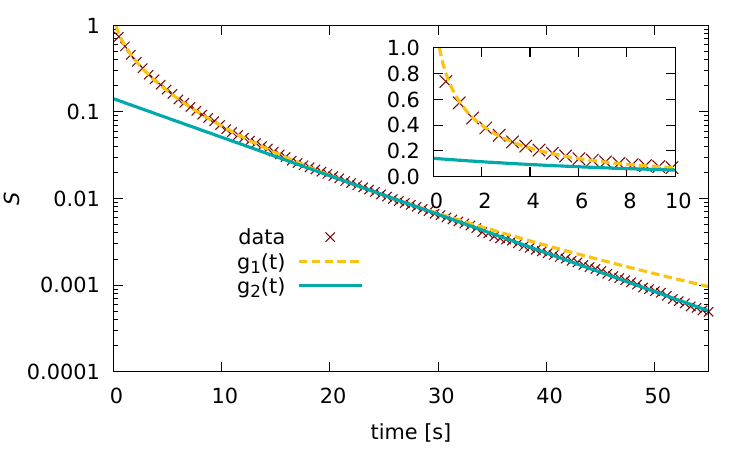}%data/hetero/swap_frac/continue2/survivors_hetero.pdf
    \caption{Fraction of surviving neurons $S(t)$ versus time in a semi-logarithmic scale.
    We fitted the data to the function $g_1(t)$ \eqref{eq_g1} in the time interval $[0:30]$ s and to the function
    $g_2(t)$ \eqref{eq_g2} in the interval $[20:55]$ s.   
    The inset shows the evolution over the first 10 seconds in a linear scale. 
    In this case we have identified $N_{\text{none}}=467$ silent neurons.
     The data refer to a system size $N=200000$ and $\sigma=0.00842$ for the clustered regime. Other parameters as in  Fig.  \ref{linear_fig}.
    }
    \label{fig_survivors_hetero}
\end{figure}

We will first examine the evolution in time of the fraction of surviving neurons $S(t)$ 
(or survival probability). As shown in Fig. \ref{fig_survivors_hetero}, $S(t)$ has an initial decay
on the interval $[0:30]$ s very well described by the following function
\begin{equation}
    g_1(t) = \alpha_1 e^{-\sqrt{t}}\,,
    \label{eq_g1}
\end{equation}
with $\alpha_1 = 1.65(1)$.  
The initial decay of $S(t)$ is followed at later times by an exponential tail of the form
\begin{equation}
    g_2(t) = \alpha_2 e^{- \beta_2 t}
    \label{eq_g2}
\end{equation}
with $\alpha_2 = 0.142(2)$ and $\beta_2 = 0.1026(3)$ Hz.
The functions $g_1$ and $g_2$ are part of the same class of survival probabilities
associated to the so-called Weibull PDF \cite{carroll2003}
\begin{equation}
f_p(t) = p \mu^p t^{p-1} {\rm e}^{-(\mu t)^p} \quad {\rm with} \quad p\in [0,+\infty) 
\quad \mu > 0 \enskip ;
\end{equation}
where $g_1$ ($g_2$) corresponds to $p=1/2$ ($p=1$). For $p <1$ the
failure rate, the rate to emit a spike in an irregular manner,
decreases over time, since the neurons that displays an
irregular spiking are eliminated from the population 
of the regular spiking ones. The neurons remaining after this initial phase
have a failure rate $\beta_2$ that is constant over time, since their survival
probability has an exponential profile, which typically emerges due to some underlying random Poissonian process.

As we will see in the following for a homogeneous network $S(t)$ is very well described by \autoref{eq_g2} over the whole time interval. Thus suggesting that this decay is likely due to the
action of the noise injected in the system, 
since this is the only source of irregularity in the homogeneous case. 
While the initial decay described by the function $g_1(t)$ should be related to the heterogeneous
distribution of the synaptic couplings. In summary, initially the neurons displaying failures in their
periodic activity are the ones with $J_i$ sufficiently different from $J_0$, while the 
successive decay involves
neurons with coupling in proximity of $J_i=J_0$. This aspect will be further analyzed in the following,
where we will correlate in more details the irregular evolution of the $i$-th neuron to its synaptic coupling
$J_i$.

\begin{figure}[htb]
    \centering
    \includegraphics[width=0.6\linewidth]{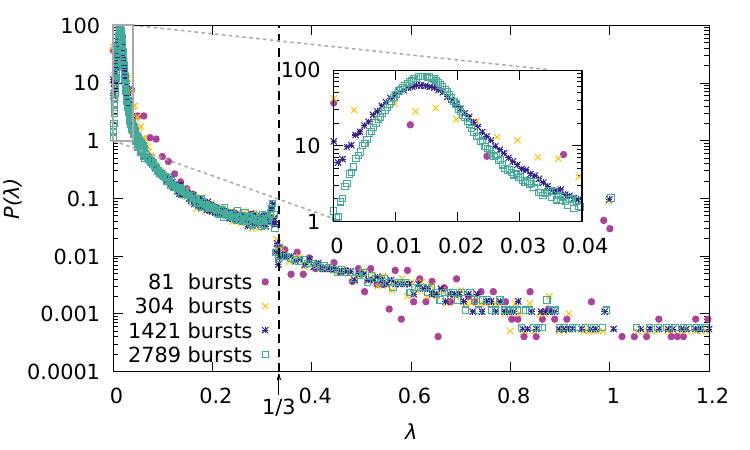}%data/hetero/swap_frac/continue2/irregular_rate.pdf
    \caption{Probability distribution function $P(\lambda)$ of the fraction of irregular spikes $\lambda$ 
    for different time durations in semi-logarithmic scale. The inset shows a zoom around the main peak.
        In this case we have removed from the estimation of the PDF $N_{\text{none}}=467$ silent neurons.
     The data refer to a system size $N=200000$ and $\sigma=0.00842$ for the clustered regime. Other parameters as in  Fig.  \ref{linear_fig}.
    }
    \label{fig_hetero_irregular_rate}
\end{figure}

Next, we examine the PDF $P(\lambda)$
of the fraction $\lambda$  of irregular spikes \eqref{eq_irregularity} 
emitted by each neuron. 
We report $P(\lambda)$ in
Fig. \ref{fig_hetero_irregular_rate} for increasing duration of the simulations and therefore
for an increasing number of population bursts.
 
From the figure it is evident that the PDF is converging to a limiting profile for
longer duration of the measurements. This asymptotic shape reveals a peak
around $\lambda \simeq 0.0145$ corresponding to the neurons
with couplings in proximity of $J_i = J_0 = -20$, i.e. to the maximum of the LD of the synaptic coyplings.
Furthermore, $P(\lambda)$ reveals a clear discontinuity at $\lambda=1/3$, whose origin will become clear in the following.

\begin{figure}[h]
    \centering
    \includegraphics[width=0.49\linewidth,scale=0.25]{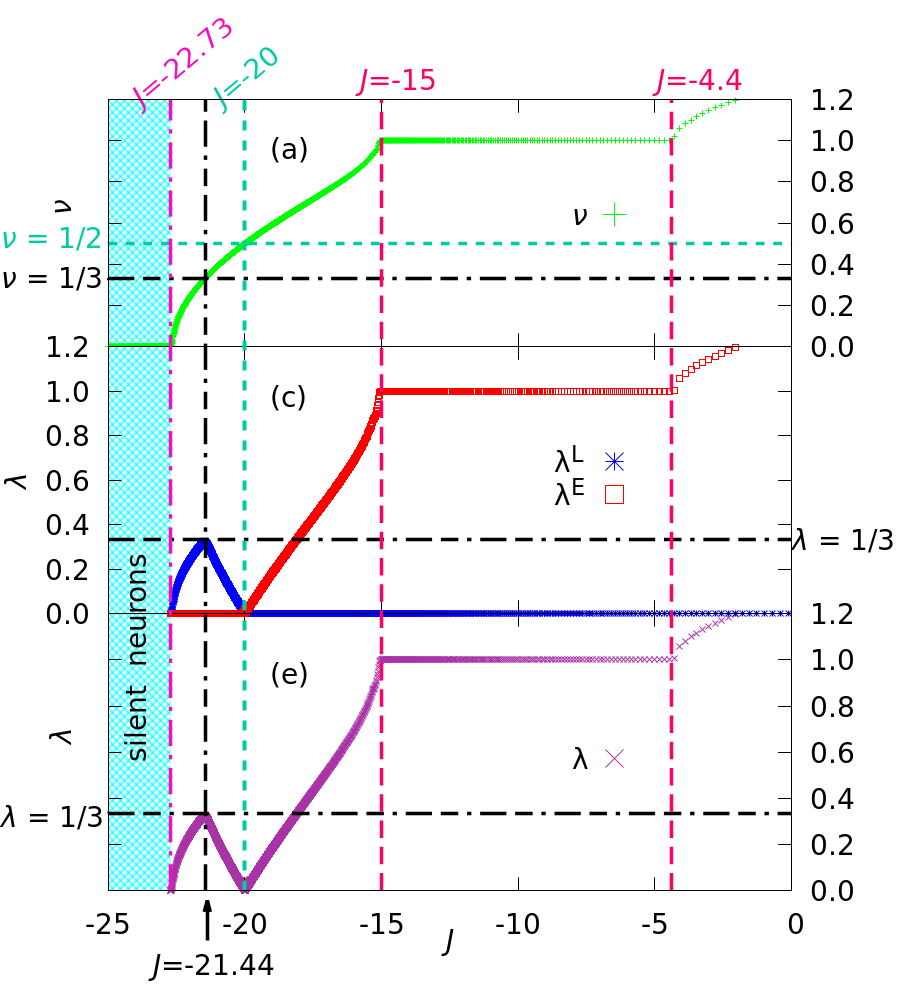}%data/hetero/swap_frac/continue2/over_j.png
  \includegraphics[width=0.49\linewidth,scale=0.25]{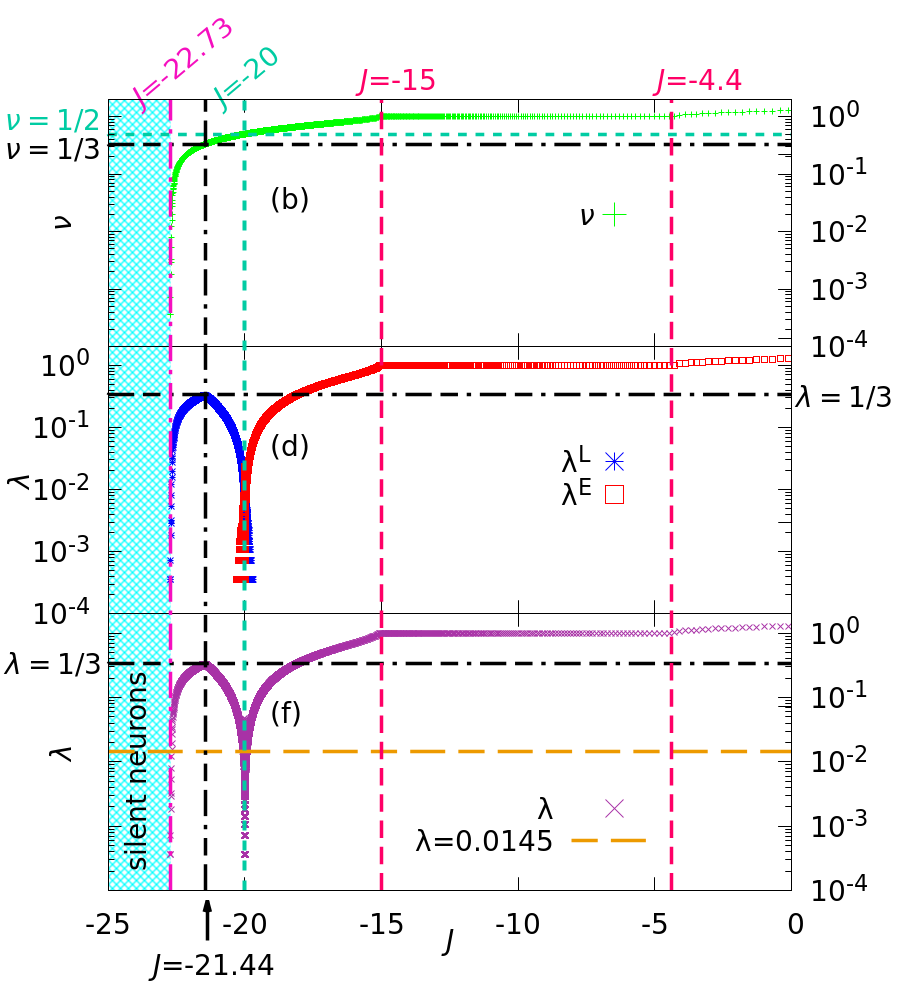}%data/hetero/swap_frac/continue2/over_j_logscale.png
    \caption{Scatter plots for $\nu_i$ (a and b) and for the indicators $\lambda_i$ (e and f), $\lambda^L_i$ and $\lambda^E_i$
    (c and d) , measuring the fraction of irregular spikes, versus the respective synaptic coupling $J_i$.
The left panels are in linear scale, while the ones on the right side are the same data reported in semi-logarithmic scale. The measurements corrrespond to the ones reported in Fig. \ref{fig_hetero_irregular_rate} for a total of 2789 bursts. }
    \label{fig_hetero_over_j}
\end{figure}

Let us now characterize in details how the 
fraction of irregularly emitted spikes of neuron $i$ depends on its synaptic coupling  $J_i$. To this aim we have estimated $\lambda_i$, $\lambda_i^E$ and $\lambda_i^L$ for each neuron as well as the 
fraction $\nu_i$ of emitted spikes  with respect to the total number of population bursts ($\nu_i$ in absence of irregularity should be $\frac{1}{2}$). These quantities are shown versus the corresponding $J_i$ as  scatter plots in Fig. \ref{fig_hetero_over_j}. 
It is important to notice that the scatter plots actually look
like smooth functions for all the considered indicators, not much ``scatter''
visible. This suggests the existence of a functional relationship between
 the measure values and the value $J$ of the synaptic coupling. 

For sufficiently small $J_i \le -22.73$ 
the neurons appear to be silent on the considered integration time
scale. As shown in panels (a and d), $\nu_i$ is growing with the synaptic coupling, apart in the locking regions. 
On the contrary, $\lambda^L_i$ and $\lambda_i$ have a non monotonic dependence on $J_i$, 
with a maximum at $J_i=-21.44$ where these parameters reach the value of exactly $\frac{1}{3}$.
$\lambda_i$ displays a minimum at $J_i=J_0=-20$, where it attains extremely small values (see panels (c and d) and (e and f)).
For larger $J_i$ essentially $\lambda_i \equiv \lambda^E_i$ and
they are both increasing with $J_i$, again apart the locking intervals. As a general remark the irregularity in the periodic firing of the neurons is due to early (late) delivered spike 
for $J_i < J_0$ ($J_i > J_0$).

Let us now try to understand the non monotonic behaviour of $\lambda$.  The local maximum  $\lambda^L_i = \lambda_i = \frac{1}{3}$ at $J_i \le -22.73$ corresponds to $\nu_i = \frac{1}{3}$,
which means that the corresponding neurons fire very regularly 
at every 3rd burst. The origin of the maximum is due to the
fact that for smaller $J_i$ the neurons are firing less and less,
thus the value of $\lambda^L_i$ and $\lambda_i$ should necessarily decrease, however for larger $J_i$ the two parameters
are also decreasing. This is due to the fact that the regular behaviour occurs whenever the neurons fire exactly every two spikes, and this state is approached by increasing $J_i$ towards
$J_0$.

Indeed, for $J_i=J_0$ the rate is  exactly $\nu_i=\frac{1}{2}$ and at the same time $\lambda_i$, $\lambda^L_i$ and $\lambda^E_i$ become quite small and close to zero (see the semi-logarithmic plots in panel (d)). In particular, from panel (f) it is evident that the neurons contributing to the peak of $P(\lambda)$ reported in Fig. \ref{fig_hetero_irregular_rate} are those with $J_i = J_0$, see the dashed orange line indicating the value $\lambda=0.0145$, where $P(\lambda)$ attains its maximum.

It is interesting to notice that in the interval  $J \in [-15,-4.4]$ we have a perfect locking of the activity of these neurons with the population bursting since $\nu=1$, and  unsurprisingly,
also $\lambda^E=1$. The locking region resembles an Arnold tongue 1:1, for $J > -4.4$ the locking is lost and $\nu$ and $\lambda$ continues to increase. We have indications that another locking region with $\nu=2$ emerges at quite large $J$, around $25\leq J \leq 29$, however our system size is too small to have a good statistics there.

Let us now come to the explanation of the discontinuity observed 
in \autoref{fig_hetero_irregular_rate} for the PDf $P(\lambda)$ at
$\lambda \approx \frac{1}{3}$. This is due to the fact that
the maximal value of $\lambda_L$ is $\frac{1}{3}$, thus for
$\lambda <  \frac{1}{3}$ to the $P(\lambda)$ contribute both
early and late delivered spikes, while for $\lambda >  \frac{1}{3}$ 
the contribution to the irregularity is due only to early delivered spikes.

To get further insight on the microscopic dynamics induced by the synaptic couplings, we define a global phase $\Phi$ similar to \autoref{phase_fire}. However, instead of the spike times of 
the individual neurons  we consider here the population burst times $b_k$:
\begin{equation}
    \Phi(t)=2 \pi \frac{t-b_k}{b_{k+1} - b_k} + 2 \pi k \quad \text{with~} b_k \leq t < b_{k+1} \enskip .
\end{equation}
Moreover, $\Phi(t)$ can be employed to characterize the activity of the $i$-th neuron with respect
to the network activity by defining the global phase difference associated to two successive spikes of neuron $i$:
\begin{equation}
    \Delta \Phi_i (n) = \Phi(t_i^{(n+1)}) - \Phi(t_i^{(n)}) \enskip .
\end{equation}

\begin{figure}[htb]
    \centering
    \includegraphics[width=0.49\linewidth,scale=0.25]{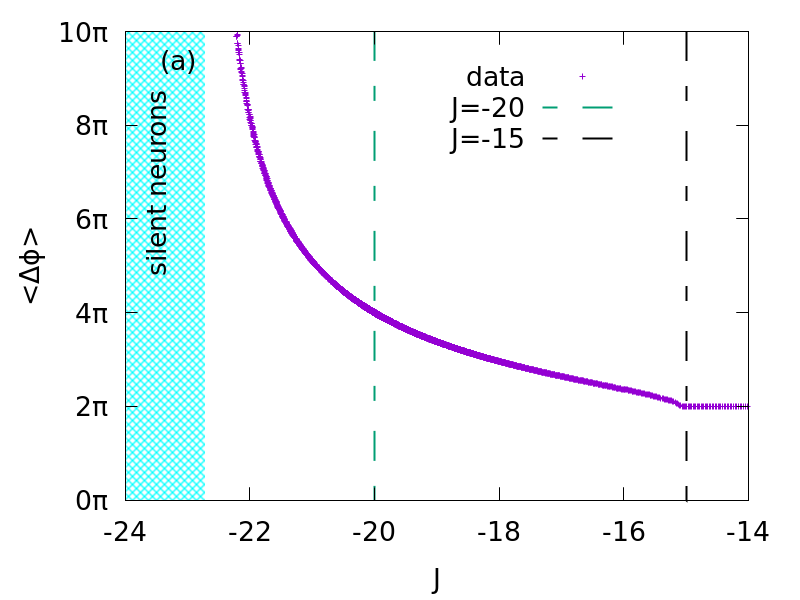}
  \includegraphics[width=0.49\linewidth,scale=0.25]{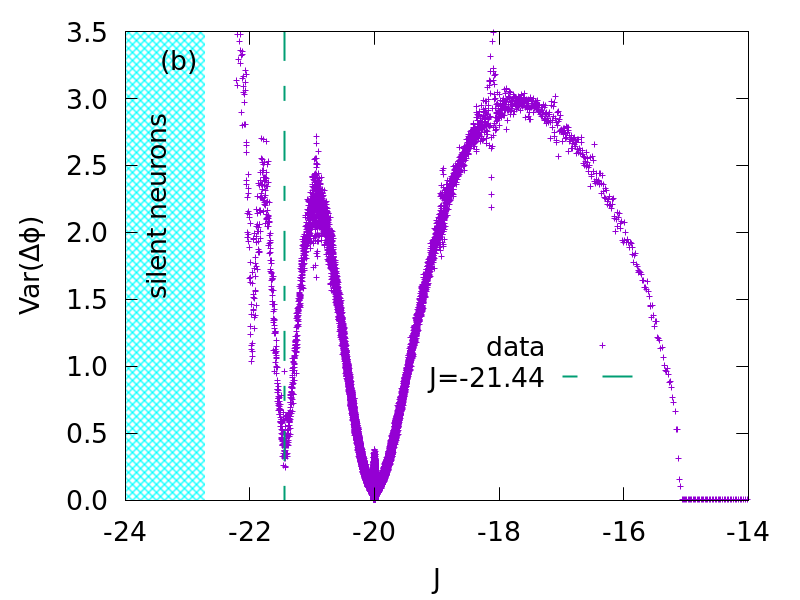}
    \caption{Mean (a) and variance (b) of the global phase difference $\Delta \Phi_i$ for each neuron
    reported as a scatter plot versus the correpsonding coupling $J_i$. The data have been obtained from the same measurements employed in Fig. \ref{fig_hetero_irregular_rate}.  
    }
    \label{global_phase_difference_fig}
\end{figure}

In \autoref{global_phase_difference_fig} we report the average and variance of
$\Delta \Phi_i$ estimated over all the spike times of neuron $i$ versus 
the corresponding synaptic coupling $J_i$. A clear functional relationship emerges
also in this case. As expected, the value of the mean of
$\Delta  \Phi_i$  is $4 \pi$  for $J_i = -20$
indicating that the neurons fire every second burst.
For $J_i < J_0$ ($J_i > J_0$) $\Delta  \Phi_i$ grows (decreases) 
indicating that the neurons fire slower (faster). Moreover,
the phase locking at $\Delta \Phi_i = 2 \pi$ for neurons with  $J_i \in [-15,-4.4]$ is also evident from panel (a).

The analysis of the variance of $\Delta \Phi_i$ reveals more interesting aspects.
 A variance close to zero suggests that the corresponding neuron fires very regularly, i.e., basically with a constant firing rate. 
In contrast a high variance indicates a distribution of the global phase
differences $\Delta \Phi_i$ exhibiting more peaks. As shown in \autoref{global_phase_difference_fig} (b), 
the variance attains its minimal value for $J_i = -J_0$
for the regular firing neurons, for which 
$\langle \Delta \Phi_i \rangle \simeq 4 \pi$. Moreover, minima
in the variance are observable also whenever $\langle \Delta \Phi_i \rangle \simeq 6 \pi$ at $J\approx -21.44$, and also at lower $J_i$ where $\langle \Delta \Phi_i \rangle \simeq 8 \pi$.
Furthermore, the variance vanishes in the locking region ($J \in [-15,-4.4]$)  where $\langle \Delta \Phi_i \rangle = 2 \pi$.
The maxima in the variance are instead observable
when $\langle \Delta \Phi_i \rangle \simeq (2 k+1) \pi$ 
for $k=0,1,2$. For the case $\langle \Delta \Phi_i \rangle \simeq 3 \pi$ ,
we observe that the corresponding neurons emit two spikes 
almost in correspondance with two successsive population bursts 
and then skip one burst. This amounts to a sequence of phase differences
$\Delta \Phi_i = 2\pi, 4\pi, 2\pi, 4\pi, \dots$, that gives an average
global phase of $3 \pi$ and a distributions of the phases with two 
equally relevant peaks and thus to a high variance.

We can safely affirm that the neurons tend to fire in correspondance of
the bursting activity of the network, in general every two bursts,
but as shown above they can present more complex combinations
of locking $n:m$ with the population bursting.

\subsection{Homogeneous synaptic couplings}

As we have seen in Subsection \ref{linear_analysis} for the case with homogeneous couplings,
i.e., $\Delta_J = \Delta_\eta = 0$, the linear stability of the mean-field model predicts that the asynchronous 
state is unstable whenever the noise amplitude is finite.  
However, the mean-field approach is no more strictly valid in the fully homogenous case, 
since the Ott-Antonsen manifold is no more attractive in such a case \cite{pikovsky2008}.
Therefore, we will limit to network simulations in order to numerically investigate 
the stability of the asynchronous state as well as possible coexistence regime 
with a collective oscillatory dynamics.

\subsubsection{The clustering transition}

\begin{figure}[htb]
    \centering
    \includegraphics[width=0.6\linewidth]{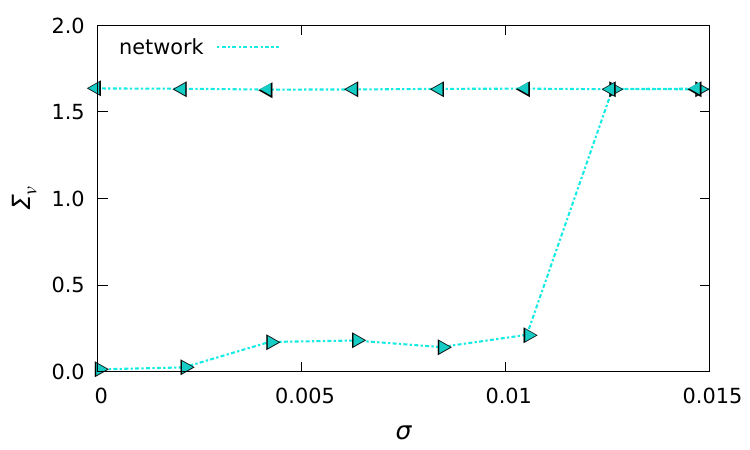}%data/homo/adiabatic/adiabatic.pdf
    \caption{
        Variance $\Sigma_v$ of the mean membrane potential $v$ versus noise amplitude $\sigma$ obtained
        via quasi-adiabatic simulations. The decrease or increase of the noise amplitude $\sigma$ performed
during the adiabatic simulations is indicated by the direction of the triangles’ tip. 
        The dashed lines are simply intended as a visual aid. The
parameters for the quasi-adiabatic simulations are  $\Delta \sigma=\frac{0.4}{19}$,  
$t_T = 20$ s, $t_S = 25$ s. The other parameters are set as in Fig. \ref{adiabatic_hetero_fig}
        with $\Delta_J=0$ for a network size $N=200000$. 
        \label{adiabatic_homo_fig}}
\end{figure}

To this aim we performed quasi-adiabatic simulations of the QIF network by varying the noise amplitude 
$\sigma$ and by evaluating the variance $\Sigma_v$ of the mean membrane potential $v$, analogously to the analysis 
done in the heterogeneous case, whose results have been reported in Fig. \ref{adiabatic_hetero_fig}.
In the present case, by increasing adiabatically $\sigma$ in the interval $[0,0.015]$ we observe 
that the asynchronous regime appears to remain stable 
up to noise amplitude $\sigma_c \simeq 0.011$, while for larger noise
COs emerge. Successively, by decreasing $\sigma$ the oscillatory regime
remains stable down to $\sigma=0$, thus we have a coexistence regime in the whole
interval $\sigma \in [0,0.011]$, as shown in Fig. \ref{adiabatic_homo_fig}.  
Note the contrast with the heterogeneous case where, in the absence of noise, only the 
asynchronous regime was observable.

The oscillatory regime is once more a clustered regime, where the neurons 
fire in population bursts and each burst involves almost half of the population. 

\subsubsection{Stability of the asynchronous regime}

\begin{figure}[htb]
    \centering
    \includegraphics[width=0.6\linewidth]{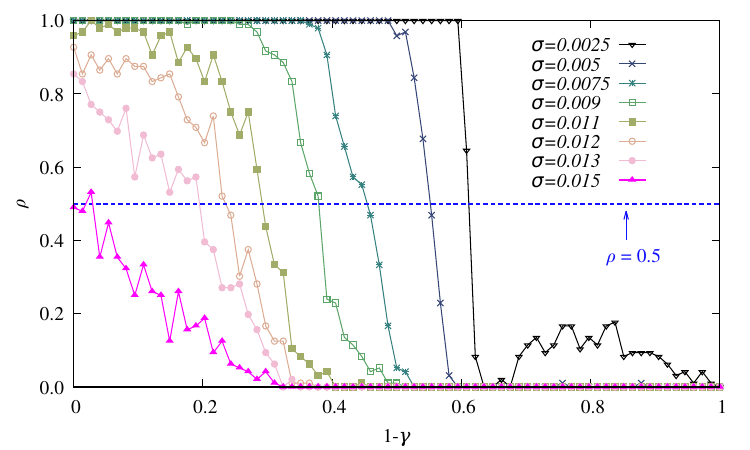}%data/homo/stability_scan/homo_stability_scan.pdf
    \caption{  
    Stability parameter $\rho$ versus $1-\gamma$, where $\gamma$ is the modulation factor entering in \eqref{v_init_eq}. 
    For each measurement of $\rho$ we averaged over $B=96$ different noise realizations and each time we estimated  $\Sigma_v$ from a time series of duration $T_W=0.3$ s after discarding a transient $T_t = 16$ s.     
    The dashed line indicates $\rho=0.5$, i.e., where the system has an equal probability to 
    end in the asynchronous or clustered state.
      The data refer to a system size $N=32000$, other parameters as in  Fig.  \ref{adiabatic_homo_fig}.}
    \label{fig_stab_homo}
\end{figure}

Let us now analyze, how the stability of the asynchronous state depends on the
noise amplitude. To perform this analysis we have employed (as in Subsection \ref{stab_asynch}) the indicator
 $\rho = \rho(\gamma)$ introduced in \autoref{stab_eq_rho}
as a function of the parameter $\gamma$. The corresponding results are reported 
in Fig. \ref{fig_stab_homo} for various noise amplitudes
$\sigma\in [0,0.015]$.

The main difference with respect to the heterogenous case, is that now even for the smallest
noise amplitude considered, namely $\sigma=0.0025$, for a sufficiently large distortion of the
LD (namely,  $\gamma > 0.4$) will destabilize the asynchronous state. This is a further confirmation
that the clustered state is always stable, as already shown in Fig. \ref{adiabatic_homo_fig}.
For increasing noise amplitudes the asynchronous state gets
destabilized for  larger and larger $\gamma$ values and for $\sigma > 0.011$ even for $\gamma=1$.
Thus indicating that this is the critical noise amplitude above which
only the clustered regime can be observed on the long-time limit.

\begin{figure}[htb]
    \centering
    \includegraphics[width=0.6\linewidth]{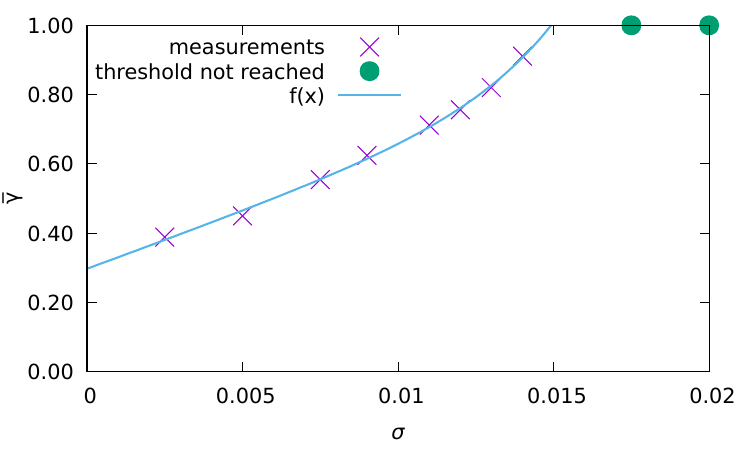}%data/homo/stability_transition/homo_transition.pdf
    \caption{${\bar \gamma}$ versus the noise amplitude $\sigma$.      
    The green circles represent $\sigma$ values for which $\rho(\sigma)=0.5$ was 
    never reached. The blue solid line is a fit to the data (violet crosses) performed via
    the expression \eqref{fit-stab-pos}, the fit is employed to extrapolate 
    the critical noise value $\sigma_{c}$ for which $\rho(\sigma_{c}, {\bar \gamma}=1)=0.5$.
    Other parameters as in Fig. \ref{fig_stab_homo} apart $B=192$. 
    }
    \label{fig_homo_gamma_trans}
\end{figure}

In analogy to what done in Subsection \ref{stab_asynch},
to better characterize this transition, we will estimate 
for various noise amplitudes the value ${\bar \gamma}$ for which 
$\rho({\bar \gamma})$ crosses $\frac{1}{2}$. 
In particular, we measured ${\bar \gamma}$ 
by using $B=192$ to obtain a better accuracy. 
The results are displayed in  Fig. \ref{fig_homo_gamma_trans}, by performing
a fit of the data to the function \eqref{fit-stab-pos} we obtained the following parameter value 
$a=-0.21(6)$, $ b= 376(205)$,  $c=33(6)$ and $\sigma_c=0.0149(3)$. Thus obtaining a critical
noise amplitude consistent with the previous estimations.

\subsubsection{Characterization of the clustered dynamics}
\begin{figure}[htb]
    \centering
    \includegraphics[width=0.6\linewidth]{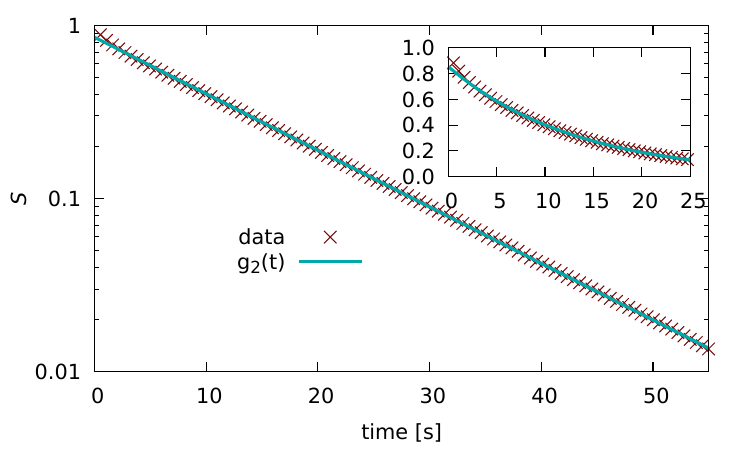}%data/homo/swap_frac/continue2/survivors_homo.pdf
    \caption{Fraction of surviving neurons $S(t)$ verus time in semi-logarithmic scale.
     We also include a fit to the numerical data with the function  $g_2(t)$ \eqref{eq_g2} (blue line).
    The inset shows the first 25 seconds of the evolution of $S(t)$ in a linear scale.
    The parameters are fixed as in Fig. \ref{adiabatic_homo_fig} for a noise amplitude $\sigma=0.00842$
    and we refer to the clustered phase.   
    }
    \label{fig_survivors_homo}
\end{figure}

Analogously to the heterogenous case the oscillatory dynamics is characterized by neurons
firing alternately in the two population bursts. 
To measure the irregularity 
in this dynamics, we have examined, as in the heterogenous case, the fraction of surviving neurons $S(t)$ defined as
in \autoref{survivor_eq}, for the same noise amplitude $\sigma=0.00842$. 
We now observe
that $S(t)$ can be well reproduced by a simple exponential decay 
\autoref{eq_g2} with parameters
$\alpha_2 = 0.853(2)$ and $\beta_2=0.0752(6)$ Hz, see Fig. \ref{fig_survivors_homo}.
The mean lifetime of the periodic regular regime is now of the order of $13.3$~s, definitely longer than in the heterogenous case. 
Since the only source of irregularity is now the noise, 
we confirm that the emergence of the irregularities in the firing process follows a Poissonian process. 
Furthermore, in contrast with the heterogeneous case we we did not observe any silent neurons, 
since none of the neurons receive very large inhibitory post-synaptic potentials, 
because the amplitude of the synaptic weight is the same for all neurons.

\begin{figure}[htb]
    \centering
    \includegraphics[width=0.6\linewidth]{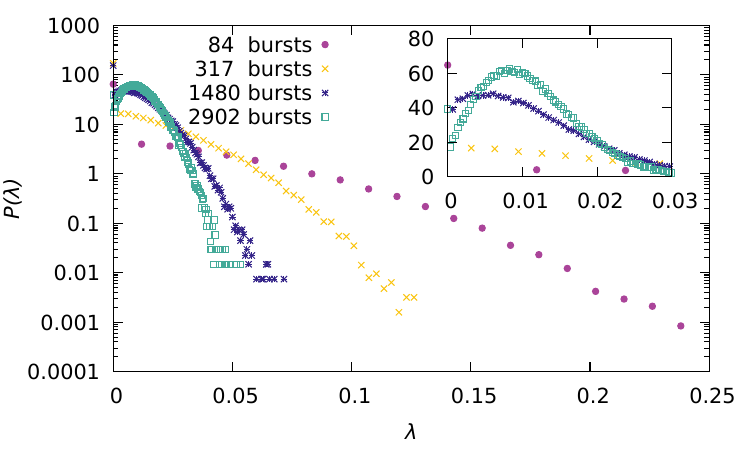}%data/homo/swap_frac/continue2/irregular_rate.pdf
    \caption{PDF $P(\lambda)$ of the fraction of irregular spikes $\lambda$ for different 
    measurement durations in semi-logarithmic scale. The inset shows a zoom in linear scale.
    The parameters are fixed as in Fig. \ref{adiabatic_homo_fig} for a noise amplitude $\sigma=0.00842$
    and we refer to the clustered phase.  
    }
    \label{fig_irregular_homo}
\end{figure}

 Next, we have estimated $P(\lambda)$, i.e. the PDF  of the fraction $\lambda$
of irregular spikes  obtained for each neuron, this is reported in Fig. \ref{fig_irregular_homo}.
We have integrated the network for the the same time 
duration as in Fig. \ref{fig_hetero_irregular_rate},
however due to the slightly higher frequency of the COs measured in the homogeneous case  
we observe more bursts.

In this case the PDF for $\lambda^E$ and $\lambda^L$ are identical (not shown) and this is
clearly due to the absence of heterogeneity in the network. The noise induces with equal probability
irregularities due to early or late spiking.

For increasing duration of the measurements we observe that $P(\lambda)$ tends to get more and more localized around the maximum located at $\lambda = \lambda_0 \approx 0.01$. Due to the central limit theorem we expect that the function  $P(\lambda)$ has a Gaussian profile (limited to $\lambda > 0$) with a standard deviation $\sigma$
scaling as $1/\sqrt{T}$, where $T$ is the time duration. Indeed, by considering only values
of $\lambda > \lambda_0$ we have verified that $\ln P(\lambda) = A + \frac{(\lambda-\lambda_0)^2}{2 \sigma^2}$
with $\sigma\propto T^\xi$ where $\xi =  -0.52 \pm 0.02$ for the data reported in Fig. \ref{fig_irregular_homo}.

\subsection{Emergence of $\gamma$-oscillations in the clustered state}

The clustered state is characterized by population bursts,
corresponding to COs.
It is of extreme interest to understand in which frequency range these oscillations occur.
In order to estimate the oscillation frequency we estimate the power spectrum density (PSD) associated to the time evolution of the mean membrane potential $v$ for heterogeneous ($\Delta_J=0.02$)
and homogeneous ($\Delta_J=0$) case subject to noise of the same amplitude, namely $\sigma=0.00842$.

\begin{figure}[htb]
    \centering
    \includegraphics[width=0.65\linewidth]{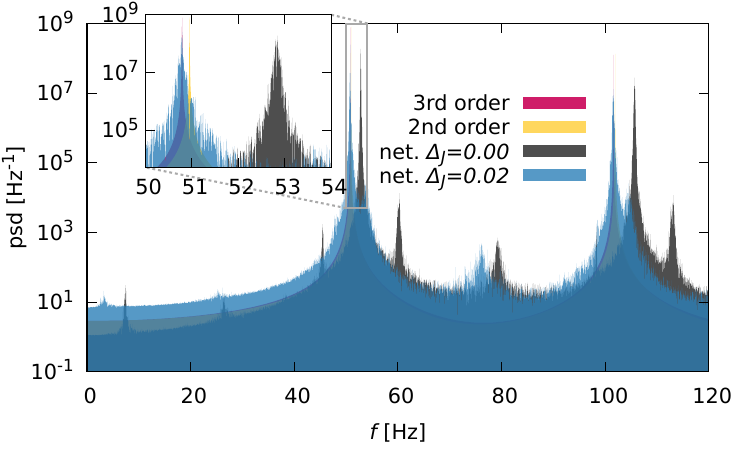}%data/hetero/fft/hetero/spectrum.pdf
    \caption{Power spectrum density (PSD) versus the the frequency $f$ of the mean membrane potential $v$ 
    for the 3rd order neural mass model (red), the 2nd order neural mass model (orange) and 
    the network simulation of size $N=200000$ with $\Delta_J=0$ (black) and $\Delta_J=0.02$ (blue), respectively.
    The colors are plotted with transparency to allow the visualization of overlapping parts. The inset shows a zoom around the global maxima.  All the other parameters are fixed as in 
 Fig. 1 and the noise amplitude to $\sigma=0.00842$.
 The PSD are measured via a discrete Fourier transform of a time series measured over 100 seconds 
with a step size of $2.5 \times 10^{-3}$ ms.
 \label{fig_spectrum_hetero}}
\end{figure}

For the heterogenous case, we observe that the two neural mass models
and the network simulations agree quite well among them as evident in  Fig.
\ref{fig_spectrum_hetero}. The only noticeable difference 
are the positions of the main peak of the PSD, that are slightly different.
The main peak of the 3rd order neural mass model and of the network
simulations are located both around $f_0 \approx 50.79~\text{Hz}$, while
the 2nd order neural mass reveals a peak located at $f_0 \approx 50.95~\text{Hz}$,
as visible in the inset of Fig. \ref{fig_spectrum_hetero}.

The simulations of the network with homogeneous couplings result in a PSD with a peak  
at a higher frequency $f_0 \approx 52.8~\text{Hz}$, still in the vicinity of 
the heterogenous peaks. 

The PSD around the peak is a bit broader for the network simulations compared to the neural mass results, since the network presents also finite-size fluctuations. In the network simulations, homogeneous and heterogeneous, we observe also peaks at
combinations of the first two harmonics, not present in the neural mass models, suggesting that finite-size effects can lead to combinations of these harmonics similar to the beating phenomenon.

\begin{figure}[htb]
    \centering
    \includegraphics[width=0.49\linewidth]{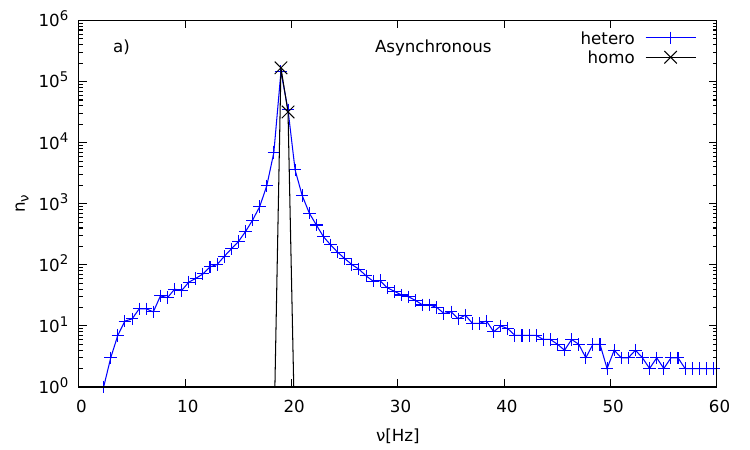}
    \includegraphics[width=0.49\linewidth]{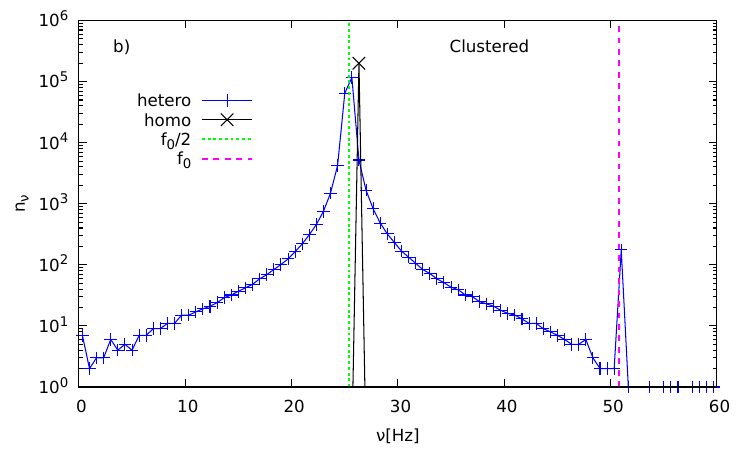}
\caption{Histogram $n_\nu$ of the single-neuron firing rate for the homogeneous and heterogeneous networks for the asynchronous (a) and clustered (b) phases. Black (red) line refers to homogeneous (heterogenous) networks. All parameters
as in Fig. \ref{fig_spectrum_hetero}.  
\label{fig_rates}}
\end{figure}

By analyzing the microscopic dynamics of the network for the same parameters in the homogenous case, we observe that the histogram $n_\nu$ 
of the single neuron firing rate $\nu_i$ is extremely localized with a peak around $\nu_0 \approx 19.27 $ Hz ($\nu_0 \approx 26.41$ Hz) for
the asynchronous (clustered) state (see Fig. \ref{fig_rates}). These data confirm that in the clustered regime the neurons mostly fire every two population bursts, 
since the frequency of COs is $f_0 \approx 52.8~\text{Hz}$. 
In the heterogenous case, the situation is more complex, as shown in  Fig. \ref{fig_rates} (b)  the histogram of the firing rates
has a main peak at $f_0/2$ with symmetric tails at lower and higher frequencies
and a secondary peak at $f_0$, where $f_0 \approx 50.79~\text{Hz}$. These data confirm the previously reported analysis for
the heterogeneous model performed in Subsection 4.1.4. 
The heterogeneity in the couplings is essentially responsible for the distributed firing rates. 
Furthermore, the most part of the neurons fire every second bursts, but a small group is locked to
the population activity.

Increasing the noise amplitude only changes the frequency slightly,
e.g., a noise of $\sigma=0.03$ results in a frequency around $52.44~\text{Hz}$ for the 3rd order mean-field neural mass, 
this means that we observe $\gamma$-oscillations in the whole region of coexistence.

\section{Summary and outlook}

We have shown that for a globally coupled inhibitory network of QIF neurons 
the presence of independent Gaussian noise promotes the emergence
of COs both for heterogeneous and homogeneous synaptic couplings.
The observed oscillations emerge at some critical noise amplitude $\sigma_c$ via a sub-critical 
Hopf bifurcations giving rise to a region of coexistence among stationary and oscillatory dynamics.
For the homogenous (heterogenous) case the coexistence is observable from zero (a finite) noise amplitude
up to $\sigma_c$. 

In the heterogenous case the analysis  is based on the comparison of the results
obtained via a direct integration of
large spiking QIF networks and of the corresponding neural mass models.
In the examined noise range we observe a quite good agreement among network simulations and
mean-field results obtained via the pseudo-cumulant expansion arrested to the third order
\cite{goldobin2021}. The second-order neural mass model fails to reproduce the
simulations for sufficiently large noise amplitudes. 

The neural mass model (at the third order) captures well the macroscopic behaviour of the
network induced by noise and heterogeneity, however being a mean-field
model cannot reproduce the microscopic dynamics, for this we should rely
on numerical simulations.

The observed collective oscillations are population bursts,
where roughly half of the neurons fire in alternation in correspondence of 
each single collective event. However, there are irregularities to this behavior.
In the heterogeneous case we observe that initially the rate at which the 
surviving neurons
emit a spike in an irregular manner decreases over time and successively
it becomes constant, i.e., it becomes a Poissonian process in the long run. 
The origin of the
initial behaviour is due to the heterogeneity in the synaptic couplings, 
while the following phase is due to the presence of the Gaussian noise.
Indeed in the homogenous case we observe only the second phase.

Furthermore, for heterogenous couplings we observe that the regular
behaviour of the neurons, i.e. firing every two population bursts,
is observable only for synaptic couplings corresponding to the median of the distribution.
For sufficiently large inhibitory couplings we have silent neurons, while
neurons displaying a 1:1 locking with the population bursts are observable
in a wide interval of synaptic couplings.
 
In order to characterize the stability of the asynchronous regime we
have introduced a new criterion based on the long-term
evolution of the system, once the stationary configuration corresponding to the asynchronous regime
has been subject to a non-infinitesimal global deformation. 
This criterion allows to identify the basins of attraction of the
two coexisting regimes, therefore resembling the  basin stability
analysis \cite{menck2013}. Furthermore, the method captures with very good accuracy 
the Hopf and saddle-node bifurcation points delimiting the coexistence regime
in the heterogenous and homogenous cases. This criterion can represent a useful
alternative to the linear stability analysis and find 
application in the context of complex networks for the characterization of their
dynamical regimes.

The nature of the noise is fundamental in order to observe the reported phenomena.
Indeed for Lorentzian distributed white noise it was shown that the corresponding
low-dimensional neural mass model \cite{cestnik2022,clusella2024} 
exhibits only a stable foci and no oscillatory regime, which we have also verified 
via network simulations.

Clustering phenomena similar to the one here analysed have been reported
in \cite{brunel2006} for globally coupled inhibitory homogenous networks for conductance based
and current based neural models in presence of Gaussian noise. However, at variance with our model the authors considered post-synaptic potentials of finite duration and not instantaneous synapses,
as in the present case. The emergence of COs 
in absence of a delay or of a finite synaptic time scale is peculiar of
inhibitory QIF networks, as previously shown in \cite{di2018}.

As we have shown for the chosen parameters the frequency of the COs is in the $\gamma$-band, 
therefore the present model can be employed to analyze the emergence of transitory $\gamma$-bursts
coexisting with asynchronous dynamics observed in many experiments  
 \cite{buzsaki2004,susin2021,spyropoulos2022,douchamps2022}. In particular, our
 model can represent a more realistic alternative to the damped 
 harmonic oscillator driven by noise employed in \cite{spyropoulos2022} to reproduce
 the emergence of spontaneous $\gamma$-cycles in awake primate visual cortex (V1).
Finally, the indicators we have introduced in Paragraph 2.3.3 to characterize the regularity/irregularity
of the single-neuron dynamics with respect to the global activity can find applications 
in the analysis of spiking events with respect to the Local Field Potential evolution
  in experimental data.

\ack
We thank D.S. Goldobin for extremely useful interactions and 
N. La Miciotta, S. Olmi, A. Pikovsky for valuable discussions
as well as Yvonne Feld for helping us in realizing several figures.
The simulations were performed at the HPC Cluster CARL, located at the University of Oldenburg
(Germany) and funded by the DFG through its Major Research Instrumentation Program (INST 184/157-1 FUGG) and the Ministry of Science and Culture (MWK) of the Lower Saxony State. 
This work also used the Scientific Compute Cluster at GWDG, the joint data center of Max Planck Society for the Advancement of Science (MPG) and University of G\"ottingen. Y.F. received financial and logistical support by the German Academic Scholarship Foundation (Studienstiftung des Deutschen Volkes). A.T. received financial support by the Labex MME-DII (Grant No. ANR-11-LBX-0023-01), by the ANR Project ERMUNDY (together with Y.F.) (Grant No. ANR-18-CE37-0014), and by CY Generations (Grant No ANR-21-EXES-0008), all part of the
French program Investissements d\'{}Avenir. Y.F. thanks the Istituto dei Sistemi Complessi (CNR) in Sesto Fiorentino, Italy
and the Laboratoire de Physique Th\'eorique et Mod\'elisation, CY Cergy Paris Universit\'e, in Cergy-Pontoise, France
for the hospitality offered during 2022 and 2023, where part of this work has been developed.

\appendix 

\section{Implementation of the numerical simulations\label{Appendix_NumericalIntegration}}

For the heterogenous case, in order to compare the results of the network simulations 
 with the neural mass models \cite{goldobin2021} we consider synaptic couplings following a LD $h(J_i)$ 
with median $J_0$ and half width at half maximum (HWHM) $\Delta_J$, that we fix deterministically as follows 
\begin{eqnarray}
    J_i = \tan \left(\frac{\pi}{2} \frac{(2 i - N - 1)}{N+1}\right) \Delta_J + J_0 \qquad \forall i \in \{1,2,\text{…}, N\}
    \quad ,
\end{eqnarray}
to avoid spurious effects related to extreme values of the couplings and to allow for a faster convergence at sufficiently large system sizes towards the corresponding mean-field results as previously shown e.g. in \cite{montbrio2015,ciszak2020}. It should be noticed that even for a LD centered at a quite negative value,
namely $J_0 = -20$ as in our case, a small number of positive coupling is expected. For the parameter values
used in this paper ($\Delta_J = 0.02 $) the percentage of excitatory synaptic coupling is 0.032 \%, therefore
we can consider their influence on the macroscopic dynamics as negligible.

The initial values of the membrane potentials are deterministically chosen as in Eq. \eqref{v_init_eq}
from the LD expected in the thermodynamic limit \eqref{LDV} for the asynchronous state. 
Note that, to avoid correlations between $V_i$ and $J_i$ that would result from the previous 
deterministic equations, the list of $J_i$ values is shuffled before creating the initial state of the network.

In order to analyze the transition from the asynchronous 
 to the partially synchronized regime due to the noise, we perform simulations where the noise amplitude  is varied quasi-adiabatically.  In particular,  we start from some initial noise value,  typically $\sigma =0$, and we simulate the models for a  certain time  interval $t_S$, after discarding a transient time $t_T$. 
 The quantities of interest are evaluated only during the interval $t_S$.  Then we increase the noise amplitude by an amount $\Delta \sigma$  and we repeat the previous procedure by an initial condition
 that is the last configuration obtained at the previous step.   The noise is increased in steps of amplitude $\Delta \sigma$ up  to some maximal value is reached.  Then the procedure is repeated by decreasing the noise at 
 each simulation step by $\Delta \sigma$ until  the initial noise value is recovered.

As already mentioned, for the QIF model the threshold value would be $V_{\rm th} =+\infty$ and the reset one $V_{\rm re}=-\infty$, it is possible to take 
in account exactly the integration among these extrema in absence of noise if the neurons are supra-threshold
by employing event driven techniques \cite{tonnelier2007,goldobin2023}. However, in presence of noise
we should perform usual clock driven simulations by 
employing finite threshold and reset values as
suggested in \cite{montbrio2015}. 

In particular, we implement the finite threshold crossing and the
spike emission as follows. Whenever $V_i(t) > V_{\rm th}$,
the neuron enter in a refractory period of duration $T_R =2/ V_i$,
after this phase the membrane potential is resetted to
$ V_i(t+T_R) = - V_i(t)$. Thus employing a variable resetting
value related to the neuron evolution, this avoids spurious
synchronization phenomena induced by using the same reset value 
for all neurons as suggested in \cite{montbrio2015}. 
Furthermore, the neuron $i$ will fire at a time $t + T_R/2$
that approximately corresponds to the time it would
reach $+\infty$ as shown in \cite{montbrio2015}.
Somehow, the usage of finite thresholds and reset value
is less matematically accurate, but it reflects more
the dynamics of real neurons \cite{koch2004}.
 
To simulate the network model we have numerically integrated 
the  stochastic differential equation \autoref{network_model}
by employing a clock driven scheme. 
In particular we have employed the Heun method \cite{SanMiguel2000}  
for its higher accuracy in the treatment of the determistic part with respect to a standard Euler scheme.

The iterative Heun method \cite{SanMiguel2000} applied to our network model reads as :
\begin{align}
    k_i &= J_i s + \Delta_t \left(\left(v_i(t)\right)^2+ \eta_i\right)\\
    l_i &= \sqrt{2 \Delta_t \sigma^2} \Xi_i \\
    v_i (t+\Delta_t/2) &= v_i(t) + l_i + k_i\\
    v_i (t+\Delta_t) &= v_i(t) + \left( \left(v_i (t+\Delta_t/2)\right)^2+\eta_i\right) \frac{\Delta_t}{2} + \frac{J_i s + k_i}{2} + l_i~,
\end{align}
where $V_i(t)$ are the membrane potentials, 
$k$ and $l$ are auxiliary variables, and 
$\Xi_i$ is a Gaussian random number with zero mean and unitary standard deviation that is drawn separately for each neuron.
Morever $\Delta_t$ is the integration time step, whose
choice will be discussed in the next Appendix, and $s$ represents
the network activity and it is the number of spikes 
emitted in the network in the interval $\Delta_t$ divided by $N$.

\section{Selection of the integration time step \label{Appendix_ChoosingStepSize}}

For the numerical integration of the network model we need to select an optimal time step $\Delta_t$, which should lead to high accuracy in the integration joined to a minimal computational cost.

In deterministic systems this choice is quite simple, 
one just select the largest time step for which the
integrated orbits converge to the same value up to some
accuracy. In a stochastic system this cannot happen,
therefore we rely on a different concept.  

In the present case, we know from the mean-field approach that in the asynchronous regime for sufficiently small noise values
$\sigma < \sigma_{SN}$ the system should always relax towards a PDF of the membrane potential that is a LD, namely
\begin{align}
p(V)=\frac{1}{\pi} \frac{\Delta_V}{\Delta_V^2+(V-V_0)^2}\,.
\label{PDF_exp}
\end{align}
with $V_0 = v^*$ and $\Delta_V = \pi r^*$, where $(v^*,r^*)$ are the fixed point solutions of the  neural mass model \autoref{neuralmass}.
Therefore, we considered a small noise value $\sigma=0.001 < \sigma_{SN}$ and for different integration time steps $\Delta_t$ 
we have verified if the distribution of the membrane potentials
converge to \eqref{PDF_exp} or not.
 
\begin{figure}[htb]
    \centering
    \includegraphics[width=0.8\linewidth]{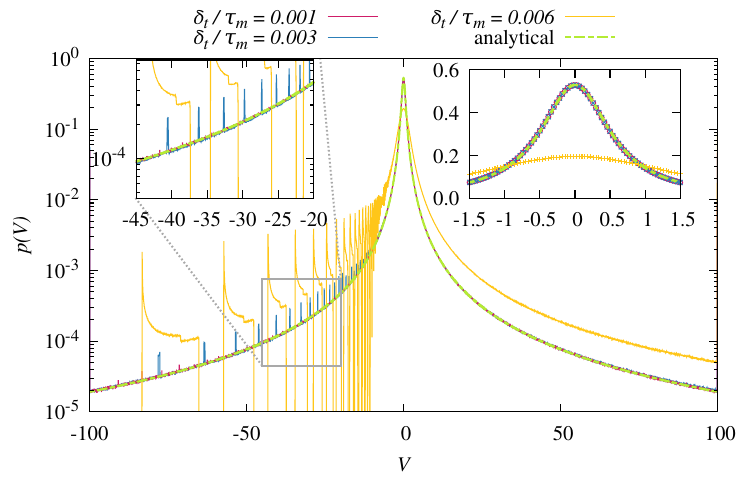}%data/hetero/step_size/sz_tmp.pdf
    \caption{Probability density function $P(v)$ as measured for a network of size $N=200000$, $\Delta J = 0.02$ and $\sigma=0.001$. 
    The dashed lime-green line shows the analytical prediction while the solid lines represent the measurements for 
    a step-size of 0.001 (red), 0.003 (blue) and 0.006 (yellow). 
    The left inset shows a zoom, while the right inset shows a range around the peak in linear scale and uses symbols for each data point that are color-coded as above.  
    }
    \label{fig_sz}
\end{figure}

In particular, we initialized the simulation
always with membrane potetials distributed as in \eqref{PDF_exp},
then we simulate the system for a time interval of 1.2 secs
and every 0.12 ms we accumulate the instantaneous values
of the membrane potentials in a histogram of 5000 bins
with   $V\in[-100,100]$. We do not consider the neurons
in their refractory periods to prevent from unphysical
overestimations of large $V$ values. 
From the final histogram we obtain the PDFs shown in
Fig. \ref{fig_sz} for three different 
$\Delta_t/\tau_m = 1 \times 10^{-3}; 3 \times 10^{-3};6 \times 10^{-3}  $
together with the expected PDF \eqref{PDF_exp}.

As evident from Fig. \ref{fig_sz}, the larger time-step 
leads to clear artifacts in the estimation of the PDF.
Already by considering $\Delta_t/\tau_m = 3 \times 10^{-3}$
leads to a noticeable improvement, in particular  
the right inset of \autoref{fig_sz} reporting $p(V)$
in linear scale around the maximum show essentially
no differences among \eqref{PDF_exp} and the estimated
PDFs with $\Delta_t/\tau_m \le 3 \times 10^{-3}$.
However, in the semi-logarithmic scale 
(left inset and main figure) the numerically estimated PDF
for $\Delta_t/\tau_m = 3 \times 10^{-3}$ still presents numerical
artifacts for sufficiently negative $V$ values.

For a time step   $\Delta_t/\tau_m = 1 \times 10^{-3}$,
we cannot notice any artifact and essentially
we have a perfect coincidence with the theoretical
PDF \eqref{PDF_exp}. We can already conclude that
this time step give a sufficient accuracy to the simulations,
however to be on the safe side we opted for
 $\Delta_t/\tau_m = 2.5 \times 10^{-4}$.

\section*{References}
\bibliographystyle{iopart-num}
\bibliography{bib}
\end{document}